\documentclass[aps,pra,twocolumn,nobibnotes]{revtex4}
\usepackage[utf8]{inputenc}
\usepackage[T1]{fontenc}
\usepackage{bm}
\usepackage{ulem}
\usepackage{epsfig}
\usepackage{graphicx}
\usepackage{amssymb,amsmath,amsbsy,amsgen,amsfonts}
\usepackage{dcolumn}
\usepackage{amsthm}
\usepackage{mathtools}
\usepackage{mathrsfs}
\usepackage{latexsym}
\usepackage{array}
\usepackage{color}
\usepackage{amstext}
\usepackage{multirow}
\usepackage{bbold}
\usepackage{color}
\usepackage{xcolor}
\usepackage{listings}
\usepackage{tikz}
\usepackage{lipsum}
\usepackage{textcomp}
\usepackage{qcircuit}

\makeatletter
\allowdisplaybreaks[1]

\makeatletter
\let\old@makecaption=\@makecaption
\usepackage{caption}
\usepackage{subcaption}
\let\@makecaption=\old@makecaption
\makeatother

\usepackage{epstopdf} 

\DeclareSymbolFont{symbols}{OMS}{cmsy}{m}{n}

\newcommand{\ket}[1]{\ensuremath{\left|#1\right>}}
\newcommand{\bra}[1]{\ensuremath{\left<#1\right|}}

\newcommand{\secref}[1]{Sec.~\ref{#1}}
\newcommand{\appendref}[1]{Appendix~\ref{#1}}
\newcommand{\equatref}[1]{Eq.~(\ref{#1})}
\newcommand{\figref}[1]{Fig.~\ref{#1}}
\newcommand{\tabref}[1]{Table~\ref{#1}}

\begin{document}

\title{Measurement-based interleaved randomised benchmarking using IBM processors}

\author{Conrad Strydom}
\email{conradstryd@gmail.com}
\author{Mark Tame}
\affiliation{Department of Physics, Stellenbosch University, Matieland 7602, South Africa}

\begin{abstract}
Quantum computers have the potential to outperform classical computers in a range of computational tasks, such as prime factorisation and unstructured searching.  However, real-world quantum computers are subject to noise.  Quantifying noise is of vital importance, since it is often the dominant factor preventing the successful realisation of advanced quantum computations.  Here we propose and demonstrate an interleaved randomised benchmarking protocol for measurement-based quantum computers that can be used to estimate the fidelity of any single-qubit measurement-based gate.  We tested the protocol on IBM superconducting quantum processors by estimating the fidelity of the Hadamard and $T$ gates --- a universal single-qubit gate set.  Measurements were performed on entangled cluster states of up to 31 qubits.  Our estimated gate fidelities show good agreement with those calculated from quantum process tomography.  By artificially increasing noise, we were able to show that our protocol detects large noise variations in different implementations of a gate.
\end{abstract}


\maketitle

\section{Introduction}\label{sec:introduction} 

Quantum computers employ the non-classical features of quantum mechanics, such as superposition and entanglement, to substantially speed up certain computational tasks such as prime factorisation~\cite{application1}, unstructured searching~\cite{application2}, simulating many body systems~\cite{application3}, machine learning~\cite{application4} and combinatorial optimisation~\cite{application5}.  In the circuit model, quantum computing is performed by explicitly applying unitary operations (or gates) from a universal set~\cite{QCM1, QCM2}.  Measurement-based quantum computing is a competing model, where the entire computation is carried out by performing adaptive single-qubit measurements on an entangled resource state~\cite{MBQC1, MBQC2, MBQC3, MBQC4}.  Hence, provided that the entangled resource state can be generated, quantum computing can be reduced to performing single-qubit measurements and multi-qubit entangling operations do not need to be performed on demand.  This is highly advantageous in linear optical systems~\cite{MBQCexam1, MBQCexam2, MBQCexam3}, where these entangling operations cannot be performed deterministically.  Measurement-based quantum computing also has benefits in physical systems such as superconducting systems~\cite{MBQCexam4}, cold atoms~\cite{MBQCexam5} and quantum dots~\cite{MBQCexam6}, where the entangled resource state is easy to generate, since the qubits to be entangled are spatially close to each other.  In our work, we focus on measurement-based quantum computing and the benchmarking of quantum gates in this model.

Irrespective of the quantum computing model or the physical system used, however, experimental realisations of quantum computers are subject to noise.  Characterising the noise is important, since noise is often the dominant factor preventing the successful realisation of advanced quantum algorithms on near-term quantum computers~\cite{noise1, noise2, noise3}.  A complete characterisation of errors on a quantum computer can be obtained by performing quantum process tomography for all its elementary operations (or hardware implemented gates) and calculating the associated gate fidelities~\cite{QPT1, QPT2, QPT3, fidelity}.  However, this is very resource intensive, since the number of experiments that need to be performed grows exponentially with the number of qubits.  This method also requires that state preparation and measurement errors are negligible, which is rarely the case in current hardware.  Randomised benchmarking, which provides a partial characterisation of errors, avoids both these problems~\cite{RBthesis}.

Standard randomised benchmarking can be used to estimate the average gate fidelity of a set of gates on a quantum computer (usually the Clifford group or a subset thereof)~\cite{SRB1, SRB2, SRB3, SRB4, SRB5, SRB6, SRB7, SRB8, SRB9, SRB10, SRB11, SRB12}.  Interleaved randomised benchmarking can be used to estimate the fidelity of individual Clifford gates~\cite{IRB1, IRB2, IRB3}.  Special interleaved randomised benchmarking protocols for estimating the fidelity of non-Clifford gates such as the $T$ gate have been proposed~\cite{IRB4, IRB5}.  Since the Clifford gates together with the $T$ gate form a universal set, these protocols enable fidelity estimation of individual gates from a universal set.

Both standard and interleaved randomised benchmarking have been implemented in a great variety of physical systems.  These include trapped ions~\cite{SRB9, IRB2, RBtrappedions1}, superconducting systems~\cite{IRB1, RBsuperconducting1, RBsuperconducting2, RBsuperconducting3, RBsuperconducting4}, nuclear magnetic resonance quantum processors~\cite{RBNMR1}, cold atoms~\cite{RBcoldatoms1, RBcoldatoms2} and quantum dots~\cite{RBqdot1}.  Several variations of randomised benchmarking have also recently been demonstrated in trapped ions~\cite{MRB} and superconducting systems~\cite{HRB, RCS}.

\begin{figure}
    \centering
    \begin{subfigure}[b]{.48\textwidth}
        \centering
        \begin{tikzpicture}[scale=0.7, font=\tiny]
        \draw[->] (7.5, 1.25) -- (7.5, 0.75);
        \node[] at (8, 1) {\footnotesize $\rho_{\text{in}}$};
        \draw[->] (3.5, 0.75) -- (3.5, 1.25);
        \node[] at (4.1, 1) {\footnotesize $\rho_{\text{out}}$};
        \filldraw[fill=gray!65, draw=black] (3, 1) circle[radius=0.25] node {6};
        \filldraw[fill=gray!65, draw=black] (7, 1) circle[radius=0.25] node {17};
        \draw (0, 0) circle[radius=0.25] node {0};
        \filldraw[fill=gray!65, draw=black] (1, 0) circle[radius=0.25] node {1};
        \filldraw[fill=gray!65, draw=black] (2, 0) circle[radius=0.25] node {4};
        \filldraw[fill=gray!65, draw=black] (3, 0) circle[radius=0.25] node {7};
        \draw (4, 0) circle[radius=0.25] node {10};
        \draw (5, 0) circle[radius=0.25] node {12};
        \draw (6, 0) circle[radius=0.25] node {15};
        \filldraw[fill=gray!65, draw=black] (7, 0) circle[radius=0.25] node {18};
        \filldraw[fill=gray!65, draw=black] (8, 0) circle[radius=0.25] node {21};
        \filldraw[fill=gray!65, draw=black] (9, 0) circle[radius=0.25] node {23};
        \filldraw[fill=gray!65, draw=black] (1, -1) circle[radius=0.25] node {2};
        \draw (5, -1) circle[radius=0.25] node {13};
        \filldraw[fill=gray!65, draw=black] (9, -1) circle[radius=0.25] node {24};
        \filldraw[fill=gray!65, draw=black] (1, -2) circle[radius=0.25] node {3};
        \filldraw[fill=gray!65, draw=black] (2, -2) circle[radius=0.25] node {5};
        \filldraw[fill=gray!65, draw=black] (3, -2) circle[radius=0.25] node {8};
        \filldraw[fill=gray!65, draw=black] (4, -2) circle[radius=0.25] node {11};
        \filldraw[fill=gray!65, draw=black] (5, -2) circle[radius=0.25] node {14};
        \filldraw[fill=gray!65, draw=black] (6, -2) circle[radius=0.25] node {16};
        \filldraw[fill=gray!65, draw=black] (7, -2) circle[radius=0.25] node {19};
        \filldraw[fill=gray!65, draw=black] (8, -2) circle[radius=0.25] node {22};
        \filldraw[fill=gray!65, draw=black] (9, -2) circle[radius=0.25] node {25};
        \draw (10, -2) circle[radius=0.25] node {26};
        \draw (3, -3) circle[radius=0.25] node {9};
        \draw (7, -3) circle[radius=0.25] node {20};
        \draw (3, 0.25) -- (3, 0.75);
        \draw (7, 0.25) -- (7, 0.75);
        \draw (0.25, 0) -- (0.75, 0);
        \draw (1.25, 0) -- (1.75, 0);
        \draw (2.25, 0) -- (2.75, 0);
        \draw (3.25, 0) -- (3.75, 0);
        \draw (4.25, 0) -- (4.75, 0);
        \draw (5.25, 0) -- (5.75, 0);
        \draw (6.25, 0) -- (6.75, 0);
        \draw (7.25, 0) -- (7.75, 0);
        \draw (8.25, 0) -- (8.75, 0);
        \draw (1, -0.25) -- (1, -0.75);
        \draw (5, -0.25) -- (5, -0.75);
        \draw (9, -0.25) -- (9, -0.75);
        \draw (1, -1.25) -- (1, -1.75);
        \draw (5, -1.25) -- (5, -1.75);
        \draw (9, -1.25) -- (9, -1.75);
        \draw (1.25, -2) -- (1.75, -2);
        \draw (2.25, -2) -- (2.75, -2);
        \draw (3.25, -2) -- (3.75, -2);
        \draw (4.25, -2) -- (4.75, -2);
        \draw (5.25, -2) -- (5.75, -2);
        \draw (6.25, -2) -- (6.75, -2);
        \draw (7.25, -2) -- (7.75, -2);
        \draw (8.25, -2) -- (8.75, -2);
        \draw (9.25, -2) -- (9.75, -2);
        \draw (3, -2.25) -- (3, -2.75);
        \draw (7, -2.25) -- (7, -2.75);
        \end{tikzpicture}
        \caption{\textit{ibm\_hanoi}}
        \label{fig:hanoitiny}
    \end{subfigure}
    
    \vspace{0.5cm}
    
    \begin{subfigure}[b]{.48\textwidth}
        \centering
        \begin{tikzpicture}[scale=0.7, font=\tiny]
        \draw[->] (6.25, 0.5) -- (5.75, 0.5);
        \node[] at (6, 0.9) {\footnotesize $\rho_{\text{in}}$};
        \draw[->] (4.25, -7.5) -- (3.75, -7.5);
        \node[] at (4, -7.1) {\footnotesize $\rho_{\text{out}}$};
        \filldraw[fill=gray!65, draw=black] (0, 0) circle[radius=0.25] node {0};
        \filldraw[fill=gray!65, draw=black] (1, 0) circle[radius=0.25] node {1};
        \filldraw[fill=gray!65, draw=black] (2, 0) circle[radius=0.25] node {2};
        \filldraw[fill=gray!65, draw=black] (3, 0) circle[radius=0.25] node {3};
        \filldraw[fill=gray!65, draw=black] (4, 0) circle[radius=0.25] node {4};
        \filldraw[fill=gray!65, draw=black] (5, 0) circle[radius=0.25] node {5};
        \filldraw[fill=gray!65, draw=black] (6, 0) circle[radius=0.25] node {6};
        \draw (7, 0) circle[radius=0.25] node {7};
        \draw (8, 0) circle[radius=0.25] node {8};
        \draw (9, 0) circle[radius=0.25] node {9};
        \filldraw[fill=gray!65, draw=black] (0, -1) circle[radius=0.25] node {10};
        \draw (4, -1) circle[radius=0.25] node {11};
        \draw (8, -1) circle[radius=0.25] node {12};
        \filldraw[fill=gray!65, draw=black] (0, -2) circle[radius=0.25] node {13};
        \filldraw[fill=gray!65, draw=black] (1, -2) circle[radius=0.25] node {14};
        \filldraw[fill=gray!65, draw=black] (2, -2) circle[radius=0.25] node {15};
        \draw (3, -2) circle[radius=0.25] node {16};
        \draw (4, -2) circle[radius=0.25] node {17};
        \draw (5, -2) circle[radius=0.25] node {18};
        \draw (6, -2) circle[radius=0.25] node {19};
        \draw (7, -2) circle[radius=0.25] node {20};
        \draw (8, -2) circle[radius=0.25] node {21};
        \draw (9, -2) circle[radius=0.25] node {22};
        \draw (10, -2) circle[radius=0.25] node {23};
        \filldraw[fill=gray!65, draw=black] (2, -3) circle[radius=0.25] node {24};
        \draw (6, -3) circle[radius=0.25] node {25};
        \draw (10, -3) circle[radius=0.25] node {26};
        \draw (0, -4) circle[radius=0.25] node {27};
        \draw (1, -4) circle[radius=0.25] node {28};
        \filldraw[fill=gray!65, draw=black] (2, -4) circle[radius=0.25] node {29};
        \filldraw[fill=gray!65, draw=black] (3, -4) circle[radius=0.25] node {30};
        \filldraw[fill=gray!65, draw=black] (4, -4) circle[radius=0.25] node {31};
        \draw (5, -4) circle[radius=0.25] node {32};
        \draw (6, -4) circle[radius=0.25] node {33};
        \draw (7, -4) circle[radius=0.25] node {34};
        \draw (8, -4) circle[radius=0.25] node {35};
        \draw (9, -4) circle[radius=0.25] node {36};
        \draw (10, -4) circle[radius=0.25] node {37};
        \draw (0, -5) circle[radius=0.25] node {38};
        \filldraw[fill=gray!65, draw=black] (4, -5) circle[radius=0.25] node {39};
        \draw (8, -5) circle[radius=0.25] node {40};
        \draw (0, -6) circle[radius=0.25] node {41};
        \draw (1, -6) circle[radius=0.25] node {42};
        \draw (2, -6) circle[radius=0.25] node {43};
        \draw (3, -6) circle[radius=0.25] node {44};
        \filldraw[fill=gray!65, draw=black] (4, -6) circle[radius=0.25] node {45};
        \filldraw[fill=gray!65, draw=black] (5, -6) circle[radius=0.25] node {46};
        \filldraw[fill=gray!65, draw=black] (6, -6) circle[radius=0.25] node {47};
        \filldraw[fill=gray!65, draw=black] (7, -6) circle[radius=0.25] node {48};
        \filldraw[fill=gray!65, draw=black] (8, -6) circle[radius=0.25] node {49};
        \filldraw[fill=gray!65, draw=black] (9, -6) circle[radius=0.25] node {50};
        \filldraw[fill=gray!65, draw=black] (10, -6) circle[radius=0.25] node {51};
        \draw (2, -7) circle[radius=0.25] node {52};
        \draw (6, -7) circle[radius=0.25] node {53};
        \filldraw[fill=gray!65, draw=black] (10, -7) circle[radius=0.25] node {54};
        \draw (1, -8) circle[radius=0.25] node {55};
        \draw (2, -8) circle[radius=0.25] node {56};
        \draw (3, -8) circle[radius=0.25] node {57};
        \filldraw[fill=gray!65, draw=black] (4, -8) circle[radius=0.25] node {58};
        \filldraw[fill=gray!65, draw=black] (5, -8) circle[radius=0.25] node {59};
        \filldraw[fill=gray!65, draw=black] (6, -8) circle[radius=0.25] node {60};
        \filldraw[fill=gray!65, draw=black] (7, -8) circle[radius=0.25] node {61};
        \filldraw[fill=gray!65, draw=black] (8, -8) circle[radius=0.25] node {62};
        \filldraw[fill=gray!65, draw=black] (9, -8) circle[radius=0.25] node {63};
        \filldraw[fill=gray!65, draw=black] (10, -8) circle[radius=0.25] node {64};
        \draw (0.25, 0) -- (0.75, 0);
        \draw (1.25, 0) -- (1.75, 0);
        \draw (2.25, 0) -- (2.75, 0);
        \draw (3.25, 0) -- (3.75, 0);
        \draw (4.25, 0) -- (4.75, 0);
        \draw (5.25, 0) -- (5.75, 0);
        \draw (6.25, 0) -- (6.75, 0);
        \draw (7.25, 0) -- (7.75, 0);
        \draw (8.25, 0) -- (8.75, 0);
        \draw (0.25, -2) -- (0.75, -2);
        \draw (1.25, -2) -- (1.75, -2);
        \draw (2.25, -2) -- (2.75, -2);
        \draw (3.25, -2) -- (3.75, -2);
        \draw (4.25, -2) -- (4.75, -2);
        \draw (5.25, -2) -- (5.75, -2);
        \draw (6.25, -2) -- (6.75, -2);
        \draw (7.25, -2) -- (7.75, -2);
        \draw (8.25, -2) -- (8.75, -2);
        \draw (9.25, -2) -- (9.75, -2);
        \draw (0.25, -4) -- (0.75, -4);
        \draw (1.25, -4) -- (1.75, -4);
        \draw (2.25, -4) -- (2.75, -4);
        \draw (3.25, -4) -- (3.75, -4);
        \draw (4.25, -4) -- (4.75, -4);
        \draw (5.25, -4) -- (5.75, -4);
        \draw (6.25, -4) -- (6.75, -4);
        \draw (7.25, -4) -- (7.75, -4);
        \draw (8.25, -4) -- (8.75, -4);
        \draw (9.25, -4) -- (9.75, -4);
        \draw (0.25, -6) -- (0.75, -6);
        \draw (1.25, -6) -- (1.75, -6);
        \draw (2.25, -6) -- (2.75, -6);
        \draw (3.25, -6) -- (3.75, -6);
        \draw (4.25, -6) -- (4.75, -6);
        \draw (5.25, -6) -- (5.75, -6);
        \draw (6.25, -6) -- (6.75, -6);
        \draw (7.25, -6) -- (7.75, -6);
        \draw (8.25, -6) -- (8.75, -6);
        \draw (9.25, -6) -- (9.75, -6);
        \draw (1.25, -8) -- (1.75, -8);
        \draw (2.25, -8) -- (2.75, -8);
        \draw (3.25, -8) -- (3.75, -8);
        \draw (4.25, -8) -- (4.75, -8);
        \draw (5.25, -8) -- (5.75, -8);
        \draw (6.25, -8) -- (6.75, -8);
        \draw (7.25, -8) -- (7.75, -8);
        \draw (8.25, -8) -- (8.75, -8);
        \draw (9.25, -8) -- (9.75, -8);
        \draw (0, -0.25) -- (0, -0.75);
        \draw (4, -0.25) -- (4, -0.75);
        \draw (8, -0.25) -- (8, -0.75);
        \draw (0, -1.25) -- (0, -1.75);
        \draw (4, -1.25) -- (4, -1.75);
        \draw (8, -1.25) -- (8, -1.75);
        \draw (2, -2.25) -- (2, -2.75);
        \draw (6, -2.25) -- (6, -2.75);
        \draw (10, -2.25) -- (10, -2.75);
        \draw (2, -3.25) -- (2, -3.75);
        \draw (6, -3.25) -- (6, -3.75);
        \draw (10, -3.25) -- (10, -3.75);
        \draw (0, -4.25) -- (0, -4.75);
        \draw (4, -4.25) -- (4, -4.75);
        \draw (8, -4.25) -- (8, -4.75);
        \draw (0, -5.25) -- (0, -5.75);
        \draw (4, -5.25) -- (4, -5.75);
        \draw (8, -5.25) -- (8, -5.75);
        \draw (2, -6.25) -- (2, -6.75);
        \draw (6, -6.25) -- (6, -6.75);
        \draw (10, -6.25) -- (10, -6.75);
        \draw (2, -7.25) -- (2, -7.75);
        \draw (6, -7.25) -- (6, -7.75);
        \draw (10, -7.25) -- (10, -7.75);
        \end{tikzpicture}
        \caption{\textit{ibmq\_brooklyn}}
        \label{fig:brooklyntiny}
    \end{subfigure}
    \caption{Qubit topologies of the processors used in our demonstration, with the qubits used shaded grey.}
\end{figure}
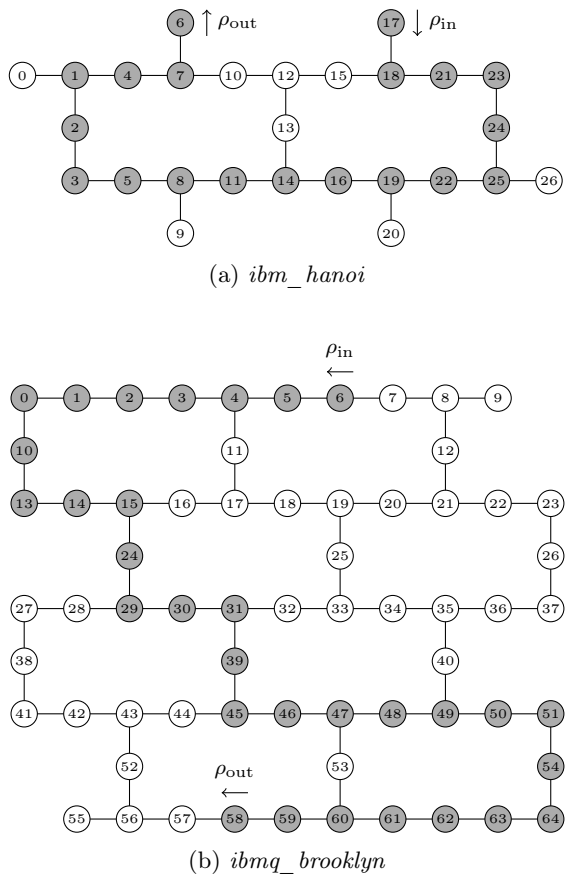

Simple adaptations of standard randomised benchmarking have been suggested for measurement-based quantum computers~\cite{MBSRB}.  In this paper we expand on the work of Ref.~\cite{MBSRB} and propose an interleaved randomised benchmarking protocol for measurement-based quantum computers.  In our measurement-based interleaved randomised benchmarking protocol, any single-qubit measurement-based 2-design can be used to estimate the fidelity of any single-qubit measurement-based gate.  We test our protocol by using the approximate measurement-based 2-design proposed and studied in our previous work~\cite{previous} to estimate the fidelity of the Hadamard gate and the $T$ gate on remotely accessible IBM superconducting quantum computers~\cite{IBM}.  Even though IBM quantum computers are primarily optimised for quantum computing in the circuit model, a variety of measurement-based protocols have been successfully implemented on these systems~\cite{noise3, previous, MBQCIBM}.  The IBM quantum computers were chosen for our experiments, since systems optimised for quantum computing in the measurement-based model, such as PsiQuantum's photonic systems, are not readily accessible and other cloud-based systems optimised for the circuit model, such as IonQ's trapped ions, had too few qubits to prepare the large entangled resource states required for our protocol at the time of starting the experiments.

Since the Hadamard gate and the $T$ gate form a universal single-qubit set~\cite{HandT}, our experiments provide a proof-of-concept demonstration of fidelity estimation of measurement-based gates from a universal single-qubit set.  In our demonstration, we use entangled resource states of up to 19 qubits on the \textit{ibm\_hanoi} quantum processor (see \figref{fig:hanoitiny}) and entangled resource states of up to 31 qubits on the \textit{ibmq\_brooklyn} quantum processor (see \figref{fig:brooklyntiny}).  In our previous work~\cite{previous}, in which we implemented single-qubit measurement-based $t$-designs on IBM processors, the implementations were not tested on any application and the entangled resource states used did not exceed six qubits.  This work therefore demonstrates significant progress in practical quantum computing on superconducting systems in the measurement-based model.

In all the experiments, estimated gate fidelities show good agreement with gate fidelities calculated from process tomography results.  Our study highlights the usefulness of IBM quantum processors for single-qubit measurement-based quantum computing and shows how to practically characterise noisy quantum logic gates in this setting.  The results of our work contribute to the ongoing goal of building up to advanced multi-qubit quantum computing on noisy intermediate scale quantum processors using the measurement-based model.

Our paper is structured as follows.  In \secref{sec:background}, we summarise single-qubit measurement-based quantum computing, single-qubit measurement-based $t$-designs and interleaved randomised benchmarking.  In \secref{sec:protocol}, we present our measurement-based interleaved randomised benchmarking protocol.  In \secref{sec:adaptation}, we discuss adjustments to the protocol that are required for implementation on the IBM quantum processors.  A description of the experiments performed and the results obtained is presented in \secref{sec:experiments}.  A summary of the work and concluding remarks are given in \secref{sec:conclusion}.  Supplementary appendices follow, in which further details about the experiments are given.

\section{Background}\label{sec:background}

\subsection{Measurement-based quantum computing}\label{sec:mbqcbackground}

In the measurement-based model, quantum computing is realised by performing adaptive single-qubit measurements on an entangled resource state such as the cluster state~\cite{MBQC1, MBQC2, MBQC3, MBQC4}.  In particular, 2D cluster states are entangled resource states for universal quantum computing~\cite{MBQC3} and fault-tolerant quantum computing can be achieved using 3D cluster states~\cite{MBQCft}.  Here we concentrate on linear cluster states, which are entangled resource states for single-qubit measurement-based quantum computing~\cite{MBQClc}.  A $n$-qubit linear cluster state is prepared on a linear array of $n$ qubits by preparing each qubit in the state $\ket{+}=\left(\ket{0}+\ket{1}\right)/\sqrt{2}$ and then applying controlled phase gates $CZ=\text{diag}(1,1,1,-1)$ between neighbouring qubits.

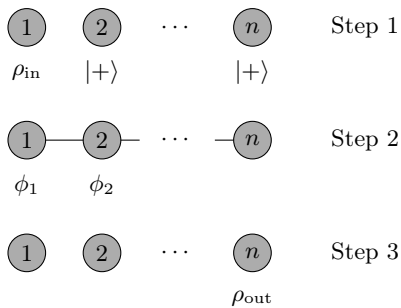
\begin{figure}
    \centering
    \begin{tikzpicture}
    \filldraw[fill=gray!65, draw=black] (0, 0) circle[radius=0.25] node {1};
    \filldraw[fill=gray!65, draw=black] (1, 0) circle[radius=0.25] node {2};
    \node[] at (2, 0) {$\cdots$};
    \filldraw[fill=gray!65, draw=black] (3, 0) circle[radius=0.25] node {$n$};
    \node[] at (4.5, 0) {Step 1};
    \node[] at (0, -0.6) {$\rho_{\text{in}}$};
    \node[] at (1, -0.6) {$\ket{+}$};
    \node[] at (3, -0.6) {$\ket{+}$};
    \filldraw[fill=gray!65, draw=black] (0, -1.5) circle[radius=0.25] node {1};
    \filldraw[fill=gray!65, draw=black] (1, -1.5) circle[radius=0.25] node {2};
    \node[] at (2, -1.5) {$\cdots$};
    \filldraw[fill=gray!65, draw=black] (3, -1.5) circle[radius=0.25] node {$n$};
    \node[] at (4.5, -1.5) {Step 2};
    \draw (0.25, -1.5) -- (0.75, -1.5);
    \draw (1.25, -1.5) -- (1.5, -1.5);
    \draw (2.5, -1.5) -- (2.75, -1.5);
    \node[] at (0, -2.1) {$\phi_{1}$};
    \node[] at (1, -2.1) {$\phi_{2}$};
    \filldraw[fill=gray!65, draw=black] (0, -3) circle[radius=0.25] node {1};
    \filldraw[fill=gray!65, draw=black] (1, -3) circle[radius=0.25] node {2};
    \node[] at (2, -3) {$\cdots$};
    \filldraw[fill=gray!65, draw=black] (3, -3) circle[radius=0.25] node {$n$};
    \node[] at (4.5, -3) {Step 3};
    \node[] at (3, -3.6) {$\rho_{\text{out}}$};
    \end{tikzpicture}
    \caption{A review of measurement-based processing using a $n$-qubit linear cluster state (adapted from Ref.~\cite{previous}).  Step~1 depicts the initialisation of the qubits.  Step~2 depicts the entangled cluster state (after controlled phase gates have been applied between neighbouring qubits) in addition to the measurements performed on each qubit.  Step~3 depicts the state resulting from the measurements.}
    \label{fig:mbp}
\end{figure}

We briefly summarise measurement-based processing with linear cluster states~\cite{previous}.  The measurement-based protocol for implementing unitary operations with a $n$-qubit linear cluster state is illustrated in \figref{fig:mbp}.  The first qubit is prepared in the input state, $\rho_{\text{in}}=\ket{\psi_{\text{in}}}\bra{\psi_{\text{in}}}$, to which the implemented unitary operation is to be applied.  The remaining qubits are prepared in the state $\ket{+}$ (Step~1), and qubits are then entangled by applying controlled phase gates between neighbouring qubits (Step~2).  Each qubit $i\in\{1,\ldots,n-1\}$ is then measured at an angle $\phi_i$ in the Pauli $XY$ plane, that is in the basis $\left\{\ket{\prescript{+}{\ }{\phi_i}},\ket{\prescript{-}{\ }{\phi_i}}\right\}$ with $\ket{\prescript{\pm}{\ }{\phi_i}}=(\ket{0}\pm e^{-\text{i}\phi_i}\ket{1})/\sqrt{2}$.  Each measurement is equivalent to applying the unitary operation
\begin{equation}
U_{m_i}(\phi_i)=X^{m_i}HR_{z}(\phi_i),
\label{eq:unitary}
\end{equation}
to $\ket{\psi_{\text{in}}}$, where $m_i\in\{0,1\}$ is the measurement outcome (occurring with probability $p_{m_i}=\frac{1}{2}$ for all $m_i$), $X$ is the Pauli $X$ operation, $H$ is the Hadamard gate and $R_{z}(\phi_i)=e^{-\text{i}Z\phi_i/2}$ is a $z$-rotation by the angle $\phi_i$.  Hence, these measurements result in the $n^{\text{th}}$ qubit being prepared in the output state, $\rho_{\text{out}}=U_{\boldsymbol{m}}(\boldsymbol{\phi})\rho_{\text{in}}U^{\dagger}_{\boldsymbol{m}}(\boldsymbol{\phi})$ (Step~3), with
\begin{equation}
U_{\boldsymbol{m}}(\boldsymbol{\phi})=U_{m_{n-1}}(\phi_{n-1})\cdots U_{m_{1}}(\phi_{1}),
\label{eq:unitarycluster}
\end{equation}
and where $\boldsymbol{\phi}$ and $\boldsymbol{m}$ denote ordered lists of angles and measurement outcomes (occurring with probability $p_{\boldsymbol{m}}=\frac{1}{2^{n-1}}$ for all $\boldsymbol{m}$) respectively.

Even though the implemented unitary operation is probabilistic and depends on the measurement outcomes, any deterministic quantum computation can be realised by employing adaptive measurement feedforward~\cite{MBQC4, MBQClc}.  In particular, unitary operations which are deterministic up to a single known Pauli gate can be implemented by performing the measurements on qubits 1 to $n-1$ sequentially and allowing future measurement angles to depend on past measurement outcomes.  The desired quantum computation is then realised deterministically by applying the required and known Pauli correction to qubit $n$.

We now give explicit measurement-based implementations of the Hadamard gate and the $T=\text{diag}(1,e^{\text{i}\pi/4})$ gate.  When all measurement outcomes are zero, the 2-qubit linear cluster state with the measurement angle $\phi_1=0$ implements the Hadamard gate and the 3-qubit linear cluster state with the measurement angles $\phi_1=\frac{\pi}{4}$ and $\phi_2=0$ implements the $T$ gate.  When non-zero measurement outcomes occur, the desired gate can still be implemented by applying the appropriate Pauli correction to the final qubit~\cite{MBQClc}.  In particular, the Pauli $X$ operation must be applied to implement the Hadamard gate when $m_1=1$.  For the $T$ gate, the Pauli $X$ operation must be applied when $m_1=0$ and $m_2=1$, the Pauli $Y$ operation must be applied when $m_1=1$ and $m_2=1$, and the Pauli $Z$ operation must be applied when $m_1=1$ and $m_2=0$.  Hence the Hadamard gate and the $T$ gate can be realised deterministically with fixed measurement angles and a Pauli correction, and in these specific cases do not require adaptive measurement feedforward.  However, implementations of more complicated deterministic operations, such as arbitrary single-qubit rotations, require both adaptive measurement feedforward and a Pauli correction~\cite{MBQC3}.

\subsection{Measurement-based t-designs}\label{sec:mbtbackground}

A unitary $t$-design is a pseudorandom ensemble of unitary operators of which the statistical moments match those of the uniform Haar ensemble either approximately or exactly up to some finite order $t$.  The expectation of any $\rho\in\mathcal{B}((\mathbb{C}^d)^{\otimes t})$ with respect to the Haar measure on the unitary group $U(d)$, where $d=2$ for single qubits, is defined by
\begin{equation}
\mathbb{E}^t_H(\rho)=\int U^{\otimes t}\rho\left(U^{\otimes t}\right)^{\dagger}\,dU.
\end{equation}
Thus formally an ensemble of unitaries $\{p_i,U_i\}$ is an exact unitary $t$-design if for all $\rho\in\mathcal{B}((\mathbb{C}^d)^{\otimes t})$
\begin{equation}
\mathbb{E}^t_H(\rho)=\sum_{i}p_iU_i^{\otimes t}\rho\left(U_i^{\otimes t}\right)^{\dagger},
\end{equation}
and $\{p_i,U_i\}$ is an $\epsilon$-approximate $t$-design if there exists an $\epsilon$ such that for all $\rho\in\mathcal{B}((\mathbb{C}^d)^{\otimes t})$
\begin{equation}
(1-\epsilon)\mathbb{E}^t_H(\rho)\leq\sum_{i}p_iU_i^{\otimes t}\rho\left(U_i^{\otimes t}\right)^{\dagger}\leq(1+\epsilon)\mathbb{E}^t_H(\rho),
\label{eq:def}
\end{equation}
where the matrix inequality $A\leq B$ holds if $B-A$ is positive semidefinite~\cite{MBT1, MBT2}.  We are primarily interested in unitary 2-designs, which are sufficient for randomised benchmarking~\cite{SRB3}.

In addition to being entangled resource states for measurement-based quantum computing (see \secref{sec:mbqcbackground}), linear cluster states can be used to implement single-qubit $t$-designs by entirely foregoing adaptive measurement feedforward and Pauli corrections, that is by fixing the measurement angles $\boldsymbol{\phi}$ and considering the ensemble of unitaries $\{p_{\boldsymbol{m}},U_{\boldsymbol{m}}(\boldsymbol{\phi})\}$ for all $\boldsymbol{m}$.  In particular, the 6-qubit linear cluster state with the measurement angles $\phi_1=0$, $\phi_2=\frac{\pi}{4}$, $\phi_3=\arccos{\sqrt{1/3}}$, $\phi_4=\frac{\pi}{4}$ and $\phi_5=0$ implements an exact 2-design~\cite{MBT1} and the 5-qubit linear cluster state with the measurement angles $\phi_1=0$, $\phi_2=\frac{\pi}{4}$, $\phi_3=\frac{\pi}{4}$ and $\phi_4=0$ implements an approximate 2-design with $\epsilon=0.5$~\cite{previous}.  In our previous work we implemented both the exact measurement-based 2-design and the approximate measurement-based 2-design on IBM processors~\cite{previous}.  Neither implementation passed our test for a 2-design under the test conditions set.  However, the test results showed that the approximate 2-design implementation more closely resembled a 2-design than the exact 2-design implementation as a result of reduced noise for the smaller 5-qubit cluster state.  This is why we use the approximate 2-design, and not the exact 2-design, in our randomised benchmarking experiments in this work (presented in \secref{sec:experiments}).  We note that the implications of performing randomised benchmarking with an $\epsilon$-approximate 2-design, as opposed to an exact 2-design, are not yet well understood theoretically.  However, numerical investigations have shown that the estimated fidelity obtained when using an exact 2-design can differ from the estimated fidelity obtained when an exact 2-design is not used~\cite{SRB10}.  We find that the results obtained with the approximate 2-design are consistent with those obtained using quantum process tomography.

\subsection{Interleaved randomised benchmarking}\label{sec:irbbackground}

Unitary 2-designs can be used in interleaved randomised benchmarking, which provides an estimate of the Haar-averaged fidelity of a noisy implementation of a unitary operation or gate~\cite{IRB1}.  The Haar-averaged fidelity of a noisy implementation of an ideal gate $G$ is defined by
\begin{equation}
F_{G}(\varepsilon,\widetilde{\varepsilon})=\int\left(\text{Tr}\left(\sqrt{\sqrt{\varepsilon(\psi)}\widetilde{\varepsilon}(\psi)\sqrt{\varepsilon(\psi)}}\right)\right)^2\,d\psi,
\end{equation}
where $\varepsilon(\psi)=G\ket{\psi}\bra{\psi}G^{\dagger}$ is the channel representing the ideal implementation of $G$ and $\widetilde{\varepsilon}(\psi)$ is the channel representing the noisy experimental implementation of $G$~\cite{MBSRB}.  The average gate error is then given by $1-F_{G}(\varepsilon,\widetilde{\varepsilon})$, which is useful for quantifying the overall reliability of the implementation.  However, for some applications, such as determining thresholds for fault-tolerant quantum computing, the worst case error is required~\cite{WCE1}.  For these applications, the average gate error can be used to obtain bounds on the worst case error~\cite{fidelity, WCE2, WCE3, WCE4}.

We briefly review the interleaved randomised benchmarking protocol proposed by Magesan \textit{et al.}~\cite{IRB1}, in which the fidelity of individual Clifford gates~\cite{Clifford} can be estimated.  However, we relax the restriction to Clifford gates and review a variation of the protocol in which any single-qubit unitary 2-design $\mathcal{U}$ can be used to estimate the fidelity of any single-qubit unitary operation or gate $G$.  Section~III~A of Ref.~\cite{IRBgeneral} explains that the interleaved method of Ref.~\cite{IRB1} holds for the variation of the protocol reviewed here.  The only benefit of the restriction to Clifford gates is that the inverse of a sequence of Clifford gates can be efficiently computed on a classical computer as a result of the Gottesman-Knill theorem~\cite{GKtheorem}.  Restricting the protocol to single-qubit gates has the same benefit, since any single-qubit system can be efficiently simulated on a classical computer~\cite{MBQClc}.  The key assumptions of the protocol are that noise from any gate is time-independent, noise from the 2-design $\mathcal{U}$ is gate-independent and noise from the inverse of any sequence of gates is independent of the sequence.

Randomised benchmarking relies heavily on a unitary 2-design's ability to transform an arbitrary noise channel into a depolarising channel~\cite{SRB4, SRB5, IRB1}, a property which was studied extensively in our previous work~\cite{previousnoise}.  In interleaved randomised benchmarking, two experiments are performed, namely an experiment to determine the reference depolarising noise parameter $p_{\text{ref}}$ and an experiment to determine the interleaved depolarising noise parameter $p_{\text{int}}$.  These parameters are then used to estimate the Haar-averaged gate fidelity using
\begin{equation}
F_{G} \approx 1 - \frac{d-1}{d}\left(1-\frac{p_{\text{int}}}{p_{\text{ref}}}\right),
\label{eq:fidelity}
\end{equation}
where $d=2$ for single qubits.

The reference depolarising noise parameter $p_{\text{ref}}$ is determined as follows:
\begin{enumerate}
\item Prepare the qubit in an arbitrary, but fixed, initial state $\rho=\ket{\psi}\bra{\psi}$.
\item Choose $m$ unitary operators uniformly at random from the 2-design $\mathcal{U}$ and apply the resulting sequence of operators, $U_{m}\cdots U_{2}U_{1}$, to the state $\rho$.
\item Compute and apply the inverse of the sequence of gates applied in step~2.
\item Measure the qubit in the basis $\{\ket{\psi}\bra{\psi},I-\ket{\psi}\bra{\psi}\}$.  This measures the probability that the initial state is unchanged by the sequence applied in step~2 followed by its inverse (known as the survival probability).
\item Repeat steps 1 to 4 for different sequences of a fixed length $m$ and average the survival probability over the different sequences to obtain the sequence fidelity $F_{\text{ref}}(m)$ for a given sequence length $m$.  Guidance on choosing the number of repetitions is given in Refs.~\cite{WCE2, RBrep}.
\item Repeat steps 1 to 5 for different sequence lengths $m$ and extract the reference depolarising noise parameter $p_{\text{ref}}$ by fitting the resulting data for $F_{\text{ref}}(m)$ versus $m$ to the exponential decay model
\begin{equation}
F_{\text{ref}}(m) = A_{\text{ref}}\,p_{\text{ref}}^{m} + B_{\text{ref}}.
\label{eq:ref}
\end{equation}
The parameters $A_{\text{ref}}$ and $B_{\text{ref}}$ absorb state preparation and measurement errors, as well as the error of the inverse applied in step~3.  The relation between the number of repetitions chosen in step~5 and confidence intervals for the extracted parameters is discussed in Ref.~\cite{WCE2}.
\end{enumerate}

The interleaved depolarising noise parameter $p_{\text{int}}$ is determined as follows:
\begin{enumerate}
\item Prepare the qubit in the same fixed initial state $\rho=\ket{\psi}\bra{\psi}$.
\item Choose $m$ unitary operators uniformly at random from the 2-design $\mathcal{U}$ and apply the interleaved sequence $GU_{m}\cdots GU_{2}GU_{1}$ to the state $\rho$.
\item Compute and apply the inverse of the sequence of gates applied in step~2.
\item Measure the qubit in the basis $\{\ket{\psi}\bra{\psi},I-\ket{\psi}\bra{\psi}\}$.  This once again measures the survival probability.
\item Repeat steps 1 to 4 for a fixed $m$ and average the survival probability over the different sequences to obtain the sequence fidelity $F_{\text{int}}(m)$.
\item Repeat steps 1 to 5 for different $m$ and extract the interleaved depolarising noise parameter $p_{\text{int}}$ by fitting the resulting data for $F_{\text{int}}(m)$ versus $m$ to the exponential decay model
\begin{equation}
F_{\text{int}}(m) = A_{\text{int}}\,p_{\text{int}}^{m} + B_{\text{int}}.
\label{eq:int}
\end{equation}
\end{enumerate}

\section{Measurement-based interleaved randomised benchmarking protocol}\label{sec:protocol}

We now present an interleaved randomised benchmarking protocol for measurement-based quantum computers (which is an extension of the standard randomised benchmarking protocol proposed for measurement-based quantum computers by Alexander \textit{et al.}~\cite{MBSRB}).  In particular, we explain how the experiments to determine the reference and interleaved depolarising noise parameters (described in \secref{sec:irbbackground}) can be implemented on a single-qubit measurement-based quantum computer.  To this end, let $\mathcal{U}$ be a single-qubit measurement-based 2-design implemented by a $(k+1)$-qubit linear cluster state with the fixed measurement angles $\boldsymbol{\phi}=(\phi_{1},\ldots,\phi_{k})$ and let $G$ be a single-qubit measurement-based gate implemented by a $(\ell+1)$-qubit linear cluster state with the measurement angles $\boldsymbol{\theta}=(\theta_{1},\ldots,\theta_{\ell})$ (possibly requiring both adaptive measurement feedforward and a Pauli correction).

On a measurement-based quantum computer, the reference depolarising noise parameter $p_{\text{ref}}$ can be determined as follows:
\begin{enumerate}
\item For a given sequence length $m$, prepare a $(mk+1)$-qubit linear cluster state.  This chooses the initial state to be the Pauli $X$ eigenstate $\rho=\ket{+}\bra{+}$, which is the natural choice for measurement-based quantum computing.
\item Repeat the measurement pattern $\boldsymbol{\phi}=(\phi_{1},\ldots,\phi_{k})$ $m$ times along the length of the cluster state.  This implements the desired random sequence, $U_{m}\cdots U_{2}U_{1}$, where the inherent randomness of the measurement-based process ensures that each $U_{i}$ is chosen uniformly at random from the 2-design $\mathcal{U}$.  Since the measurement angles are fixed, the $mk$ measurements can be performed simultaneously.  The measurement outcomes can be used to determine which sequence of gates was implemented.
\item Compute the inverse of the sequence of gates implemented in step~2 and apply this inverse by performing a measurement basis rotation on the final qubit.  The inverse is applied in this way, and not as a measurement-based operation, to ensure that noise from the inverse is independent of the sequence.
\item Measure the final qubit in the Pauli $X$ basis, $\{\ket{+},\ket{-}\}$.
\item Repeat steps 1 to 4 for a fixed sequence length $m$ and determine $F_{\text{ref}}(m)$.
\item Repeat steps 1 to 5 for different sequence lengths $m$ and fit the resulting data for $F_{\text{ref}}(m)$ versus $m$ to \equatref{eq:ref}.
\end{enumerate}

On a measurement-based quantum computer, the interleaved depolarising noise parameter $p_{\text{int}}$ can be determined as follows:
\begin{enumerate}
\item For a given $m$, prepare a $(m(k+\ell)+1)$-qubit linear cluster state.
\item Repeat the measurements $(\phi_{1},\ldots,\phi_{k},\theta_{1},\ldots,\theta_{\ell})$ $m$ times along the length of the cluster state.  This implements the desired interleaved sequence, $GU_{m}\cdots GU_{2}GU_{1}$.  Since adaptive measurement feedforward may be required for the implementation of $G$, the $m(k+\ell)$ measurements need to be performed sequentially.  If Pauli corrections are required, these can be applied after each set of $k+\ell$ measurements, by performing a measurement basis rotation, before proceeding with the next set of $k+\ell$ measurements.
\item Compute the inverse of the sequence of gates implemented in step~2 and apply this inverse by performing a measurement basis rotation on the final qubit.
\item Measure the final qubit in the Pauli $X$ basis, $\{\ket{+},\ket{-}\}$.
\item Repeat steps 1 to 4 for a fixed $m$ and determine $F_{\text{int}}(m)$.
\item Repeat steps 1 to 5 for different $m$ and fit the resulting data for $F_{\text{int}}(m)$ versus $m$ to \equatref{eq:int}.
\end{enumerate}

\section{Adjustments to protocol for implementation on IBM processors}\label{sec:adaptation}

Since IBM quantum processors did not support dynamic circuit execution at the time of performing the experiments, our measurement-based interleaved randomised benchmarking protocol had to be adjusted to enable implementation on the IBM hardware available at the time.  Without dynamic circuit execution, it is not possible to include the inverse (step~3 of the experiments to determine the depolarising noise parameters) in the required quantum circuits, since the inverse depends on the sequence being implemented in step~2, which is only known once the measurements have been performed.  One solution is to construct a different circuit for each possible inverse, run each circuit a sufficient number of times to obtain the random sequence corresponding to the implemented inverse, and then use post-selection to eliminate runs in which this desired sequence was not obtained.  Since there are $2^n$ possible inverses for the sequences implemented by performing measurements on a $(n+1)$-qubit linear cluster state, we would need $2^n$ different circuits, and since each random sequence occurs with probability $\frac{1}{2^{n}}$, all but 1 in every $2^n$ runs would be eliminated, and each of the $2^n$ circuits would need to be run at least $2^n$ times to have a reasonable chance of obtaining the sequence corresponding to the implemented inverse.  Hence the number of runs required for post-selection scales like $2^{2n}$, which is generally prohibitively expensive.

We therefore employ an alternative strategy in our randomised benchmarking experiments in \secref{sec:experiments}.  We perform full quantum state tomography on the final qubit (after performing the measurements in step~2 of the experiments to determine the depolarising noise parameters) for each of the possible measurement outcomes, apply the correct inverse for each implemented sequence to the appropriate constructed density matrix by performing matrix multiplication, and then extract the survival probability from each resulting density matrix.  Since each of the $2^n$ random sequences implemented by performing measurements on a $(n+1)$-qubit linear cluster state occurs with probability $\frac{1}{2^{n}}$, the associated circuit must be run $3(500)(2^n)$ times to obtain 500 data points for tomography (which requires 3 basis measurements) for each of the $2^n$ possible measurement outcomes.  Hence the circuit must be run $3(500)(2^n)/8000=3(2^n)/16$ times if each run has 8000 shots.  In the experiments in \secref{sec:experiments}, 8192 shots (the maximum number of allowed shots on IBM quantum processors at the time of starting the experiments) were used for each run, instead of just 8000 shots, to increase the likelihood of obtaining at least 500 data points for tomography even when the measurement outcomes are not uniform as a result of noise.  Although this method has the disadvantage that the inverse is performed through classical post-processing, and not as a quantum mechanical operation, it scales like $2^n$, which is much better than post-selection, which scales like $2^{2n}$.  Since we assume that noise from the inverse is independent of the sequence, noise from the inverse would in any case not affect the depolarising noise parameters which are used to estimate the fidelity, and so performing the inverse through classical post-processing does not affect the results.

Since adaptive measurement feedforward cannot be implemented without dynamic circuit execution either, our protocol could only be used to estimate the fidelity of measurement-based gates which can be implemented with fixed measurement angles and a Pauli correction (such as the 2-qubit Hadamard gate and the 3-qubit $T$ gate given in \secref{sec:mbqcbackground}) on the IBM quantum hardware available at the time of performing the experiments.  As the Pauli corrections required in the interleaved sequences depend on the measurement outcomes, these also cannot be performed without dynamic circuit execution.  We therefore average the survival probability, not only over the different random unitaries, but also over the different byproducts that result from omitting the Pauli corrections, when calculating the interleaved sequence fidelity in the experiments in \secref{sec:experiments}.  When applying the inverse of an interleaved sequence with matrix multiplication, the inverse of each individual byproduct is applied, as determined by the measurement outcomes.

\section{Experiments}\label{sec:experiments}

\subsection{Implementation of protocol}\label{sec:expimplementation}

\begin{figure}
\centering
\hspace{0.5cm}
\Qcircuit @C=1em @R=1em {
\lstick{\ket{0}} & \gate{\text{H}} & \ctrl{1} & \qw & \qw & \qw & \gate{\text{H}} & \meter & \qw \\
\lstick{\ket{0}} & \qw & \targ & \targ & \gate{\text{H}} & \gate{\text{R}_z(\phi)} & \gate{\text{H}} & \meter & \qw \\
\lstick{\ket{0}} & \gate{\text{H}} & \ctrl{1} & \ctrl{-1} & \qw & \gate{\text{R}_z(\phi)} & \gate{\text{H}} & \meter & \qw \\
\lstick{\ket{0}} & \qw & \targ & \targ & \gate{\text{H}} & \qw & \gate{\text{H}} & \meter & \qw \\
\lstick{\ket{0}} & \gate{\text{H}} & \ctrl{1} & \ctrl{-1} & \qw & \gate{\text{R}_z(\phi)} & \gate{\text{H}} & \meter & \qw \\
\lstick{\ket{0}} & \qw & \targ & \targ & \gate{\text{H}} & \qw & \gate{\text{H}} & \meter & \qw \\
\lstick{\ket{0}} & \gate{\text{H}} & \qw & \ctrl{-1} & \qw & \qw & \qw & \gate{\text{out}} & \qw
}
\caption{Quantum circuit for the implementation of the interleaved sequence for the 3-qubit $T$ gate with $m=1$ on the \textit{ibm\_hanoi} quantum processor.  The angle $\phi=\frac{\pi}{4}$ and `out' represents the different basis measurements used to perform quantum state tomography on the final qubit.}
\label{fig:circuit}
\end{figure}
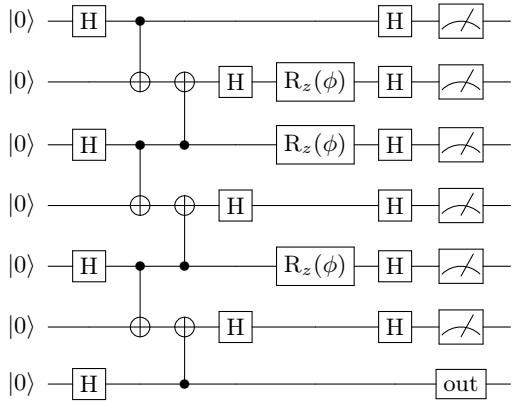

We implemented our measurement-based interleaved randomised benchmarking protocol, with the adjustments discussed in \secref{sec:adaptation}, first on the \textit{ibm\_hanoi} quantum processor, a 27-qubit superconducting IBM quantum computer.  Details of the qubits used can be found in \appendref{append:qubitsuniversal}.  We used the 5-qubit approximate measurement-based 2-design proposed in our previous work~\cite{previous} (see \secref{sec:mbtbackground}) to estimate the fidelity of the 2-qubit measurement-based implementation of the Hadamard gate and the 3-qubit measurement-based implementation of the $T$ gate given in \secref{sec:mbqcbackground} on the \textit{ibm\_hanoi} quantum processor.  To this end, we implemented reference sequences with lengths $m\in\{1,2,3\}$ and interleaved sequences with $m\in\{1,2,3\}$.  For each $m$ (reference or interleaved), we prepared the required linear cluster state, performed the appropriate single-qubit measurements on all but the final qubit, and then performed full quantum state tomography on the final qubit for each of the possible measurement outcomes to infer the output state for each corresponding random sequence.  As an example, the quantum circuit for the implementation of the interleaved sequence for the 3-qubit $T$ gate with $m=1$, which requires a 7-qubit linear cluster state, is shown in \figref{fig:circuit}.  The largest number of qubits was used for the implementation of the interleaved sequence for the 3-qubit $T$ gate with $m=3$, which required $(m(k+\ell)+1)=(3(4+2)+1)=19$ qubits, as shown in \figref{fig:hanoitiny}.

Since controlled phase gates are not supported at the hardware level on IBM quantum processors, linear cluster states were prepared using Hadamards and controlled not gates, instead of Hadamards and controlled phase gates, as recommended by Mooney \textit{et al.}~\cite{lcspreparation}.  This prevented redundant Hadamards, which would have unnecessarily increased noise, from being introduced during transpilation.  Since qubits can only be measured in the computational basis, $\{\ket{0}, \ket{1}\}$, on IBM processors, single-qubit measurements at an angle $\phi$ in the Pauli $XY$ plane were carried out by applying $R_z(\phi)$, followed by a Hadamard, and measuring the qubit in the computational basis.

To obtain the data required to determine the sequence fidelity ($F_{\text{ref}}(m)$ or $F_{\text{int}}(m)$) for a given $m$, the relevant circuit was run $3(2^n)/16$ times with 8192 shots (see \secref{sec:adaptation}) on the \textit{ibm\_hanoi} quantum processor (where $n$ is the number of single-qubit measurements performed on the $(n+1)$-qubit linear cluster state in the implementation).  The data (counts) obtained in repeated runs were combined and grouped according to measurement outcomes.  The output density matrix for each random sequence was then constructed from the tomography data obtained for the corresponding set of measurement outcomes.  To ensure that the density matrices constructed from the data are physical (i.e. that they are Hermitian and have a trace of 1) we employed qiskit's built-in method~\cite{tomography}, which uses maximum-likelihood estimation to find the closest physical density matrix to a density matrix constructed from tomography data.  The appropriate inverse was then applied to each density matrix and the survival probability was extracted from each resulting density matrix.  The sequence fidelity ($F_{\text{ref}}(m)$ or $F_{\text{int}}(m)$) presented for a given $m$ in \secref{sec:expuniversal} is the average of these survival probabilities, and the error is given by the standard deviation.

In interleaved randomised benchmarking experiments, sequence fidelities are typically determined for sequence lengths up to $m=80$ or longer~\cite{IRB1, RBsuperconducting3}.  Here, the resources required to implement longer sequences prevented us from considering sequence lengths beyond $m=3$.  For example, the implementation of the interleaved sequence for the 3-qubit $T$ gate with $m=3$ required a 19-qubit linear cluster state and the relevant circuit had to be run $3(2^{18})/16=49152$ times to obtain the data required to determine the sequence fidelity.  Since this is already extremely resource intensive, longer sequences would not be feasible, as the number of runs required grows exponentially with the length of the sequence.  Hence, even though the adjustments discussed in \secref{sec:adaptation} have enabled us to obtain data for a proof-of-concept demonstration, dynamic circuit execution remains essential for the efficient scaling of our measurement-based interleaved randomised benchmarking protocol in future.

\subsection{Results for universal gates}\label{sec:expuniversal}

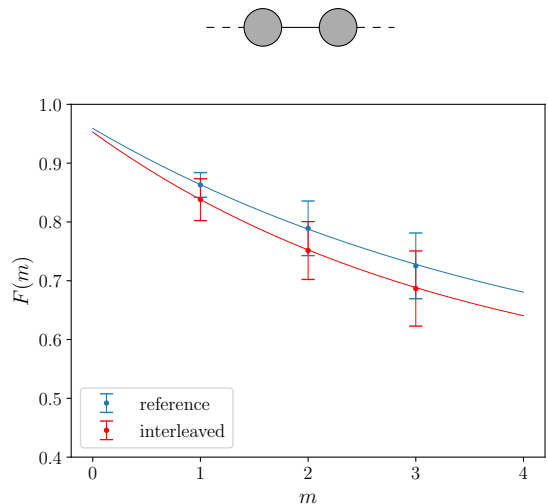
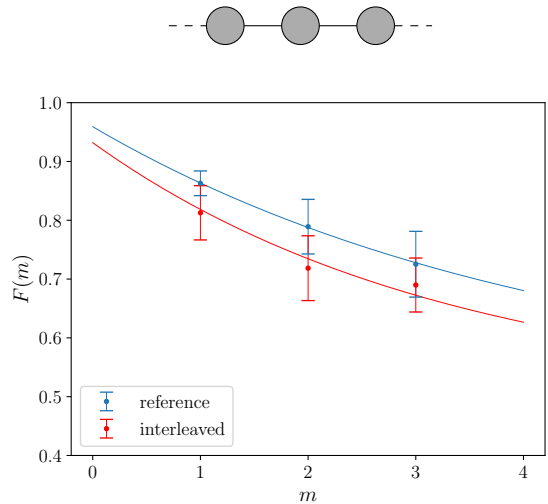
\begin{figure}
    \centering
    \begin{subfigure}[b]{.48\textwidth}
        \centering
        \begin{tikzpicture}
        \filldraw[fill=gray!65, draw=black] (0, 0) circle[radius=0.25];
        \filldraw[fill=gray!65, draw=black] (1, 0) circle[radius=0.25];
        \draw[dashed] (-0.75, 0) -- (-0.25, 0);
        \draw (0.25, 0) -- (0.75, 0);
        \draw[dashed] (1.25, 0) -- (1.75, 0);
        \end{tikzpicture}
        \includegraphics[scale=0.5]{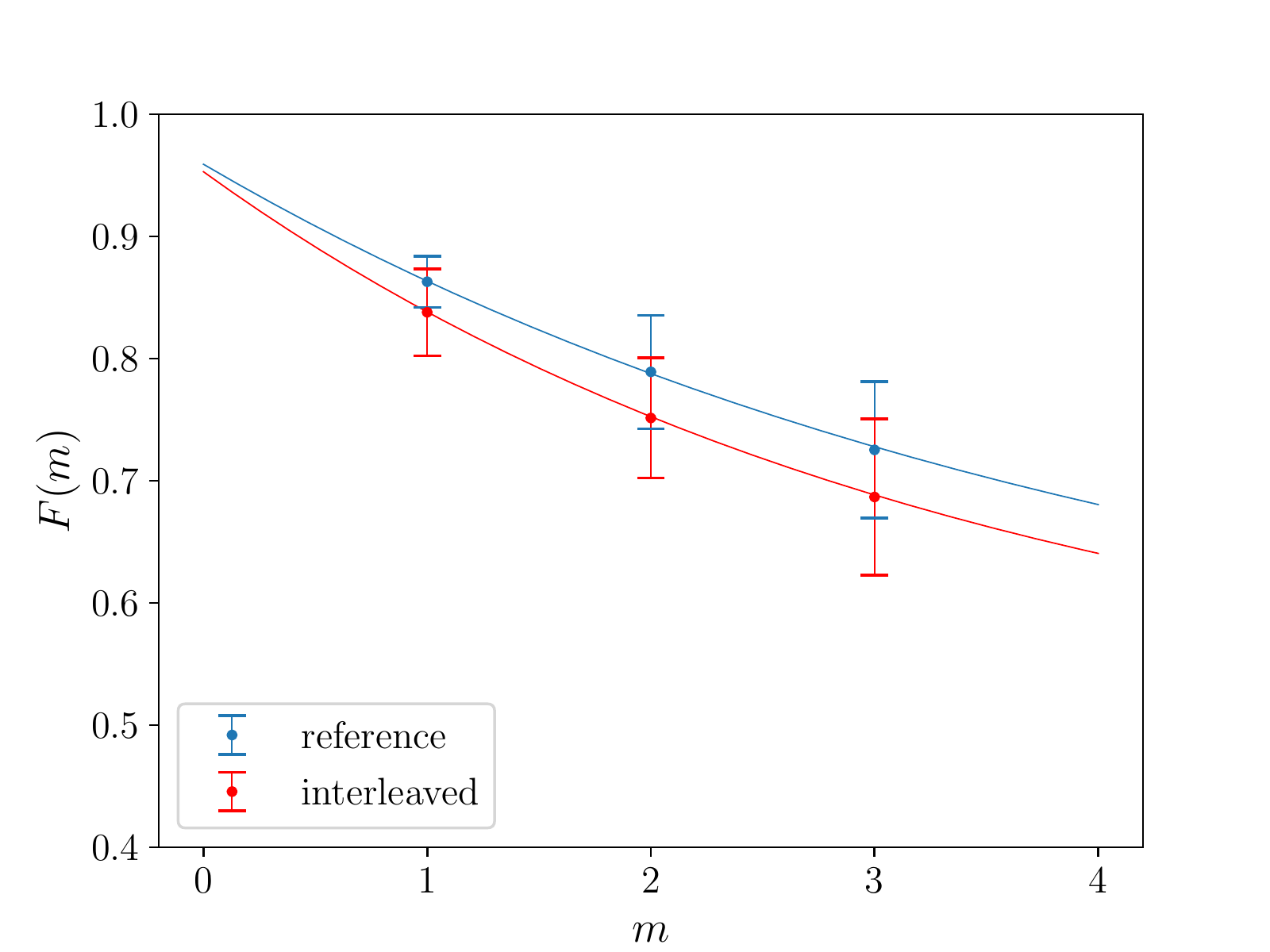}
        \caption{Hadamard gate}
    \end{subfigure}
    
    \vspace{0.75cm}
    
    \begin{subfigure}[b]{.48\textwidth}
        \centering
        \begin{tikzpicture}
        \filldraw[fill=gray!65, draw=black] (0, 0) circle[radius=0.25];
        \filldraw[fill=gray!65, draw=black] (1, 0) circle[radius=0.25];
        \filldraw[fill=gray!65, draw=black] (2, 0) circle[radius=0.25];
        \draw[dashed] (-0.75, 0) -- (-0.25, 0);
        \draw (0.25, 0) -- (0.75, 0);
        \draw (1.25, 0) -- (1.75, 0);
        \draw[dashed] (2.25, 0) -- (2.75, 0);
        \end{tikzpicture}
        \includegraphics[scale=0.5]{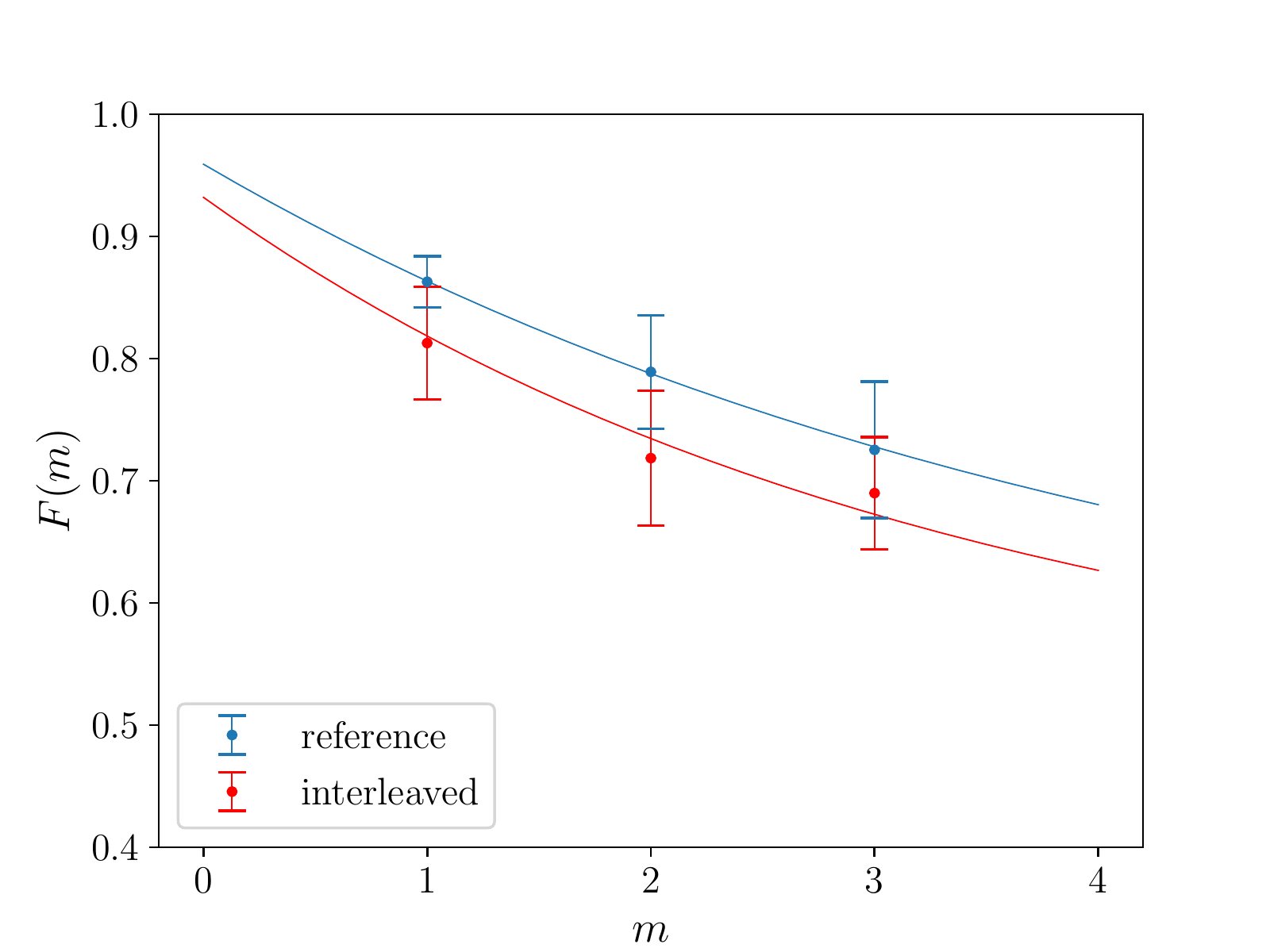}
        \caption{$T$ gate}
    \end{subfigure}
    \caption{Sequence fidelities $F(m)$ obtained for $m\in\{1,2,3\}$ to estimate the fidelity of the 2-qubit measurement-based implementation of the Hadamard gate and the 3-qubit measurement-based implementation of the $T$ gate, using the 5-qubit approximate measurement-based 2-design of Ref.~\cite{previous}, on the \textit{ibm\_hanoi} quantum processor.  Reference and interleaved sequence fidelities were fit to Eqs.~(\ref{eq:ref}) and (\ref{eq:int}) respectively (see \appendref{append:fitting}).  (a)~Hadamard gate shows the reference sequence fidelities (in blue) and the interleaved sequence fidelities for the 2-qubit Hadamard gate (in red).  (b)~$T$~gate shows the same reference sequence fidelities (in blue) and the interleaved sequence fidelities for the 3-qubit $T$ gate (in red).}
    \label{fig:universal}
\end{figure}

Sequence fidelities obtained to estimate the fidelity of the 2-qubit measurement-based implementation of the Hadamard gate and the 3-qubit measurement-based implementation of the $T$ gate (a universal single-qubit set), using the 5-qubit approximate measurement-based 2-design of Ref.~\cite{previous}, on the \textit{ibm\_hanoi} quantum processor are shown in \figref{fig:universal}.  The large uncertainties in the sequence fidelities reflect the gate-dependence of noise from the approximate measurement-based 2-design~\cite{previous}.  A Monte Carlo method which takes these uncertainties into account was used to fit the reference and interleaved sequence fidelities to the exponential decay model given by Eqs.~(\ref{eq:ref}) and (\ref{eq:int}) respectively.  The fitting procedure used, the fitting constraints imposed and the estimation of Haar-averaged gate fidelities from the fitting parameters are discussed in \appendref{append:fitting}.  The estimated fidelity of the 2-qubit measurement-based implementation of the Hadamard gate is (0.977$\pm$0.073) and the estimated fidelity of the 3-qubit measurement-based implementation of the $T$ gate is (0.972$\pm$0.070).  The large uncertainties in the estimated fidelities can be attributed both to the large uncertainties in the sequence fidelities and to the small number of sequence fidelities.  It is unclear to what extent the use of an $\epsilon$-approximate 2-design, as opposed to an exact 2-design, contributed to the large uncertainties in the estimated fidelities.  It is also not clear how $\epsilon$ is related to these uncertainties or how the relation is affected by the Monte Carlo method used to estimate the fidelities.

We now compare the fidelities estimated using our measurement-based interleaved randomised benchmarking protocol to fidelities calculated from process tomography results.  To this end, we performed quantum process tomography on each of the three respective sets of qubits of the \textit{ibm\_hanoi} quantum processor on which the 2-qubit Hadamard gate and the 3-qubit $T$ gate were implemented in the interleaved sequences, and used the results to calculate the Haar-averaged gate fidelity of each of the three implementations of the 2-qubit Hadamard gate and the 3-qubit $T$ gate.  Further details about the process tomography method used, the implementation of the method and the calculation of Haar-averaged gate fidelities from process tomography results are given in \appendref{append:channeltomo}.  Fidelities obtained for the three different implementations of the 2-qubit Hadamard gate, on the three different sets of qubits used in the interleaved sequences, range from 0.948 to 0.972.  For the 3-qubit $T$ gate, the fidelities obtained from process tomography results range from 0.928 to 0.939.

For both gates, the uncertainty of the gate fidelity calculated using process tomography results is smaller than the uncertainty of the gate fidelity estimated using our measurement-based interleaved randomised benchmarking protocol.  We also note that the estimated fidelities are slightly larger than those obtained using process tomography results.  Nevertheless, the agreement between our estimated gate fidelities and the gate fidelities calculated from process tomography results is somewhat remarkable considering that we used a very weak approximate 2-design~\cite{previous}, the implementation of this 2-design on the IBM quantum processors did not even pass our test for an approximate 2-design for all states~\cite{previous}, noise from this 2-design is not entirely gate-independent~\cite{previous} and the fitting parameters were inferred from very few sequence fidelities (only three).  This clearly demonstrates the robustness of interleaved randomised benchmarking.  It also provides motivation for considering weak approximate 2-designs, such as the one proposed in our previous work~\cite{previous}.  It is unclear to what extent the use of a weak $\epsilon$-approximate 2-design, as opposed to an exact 2-design, contributed to the over-estimation of gate fidelities.  However, one would expect that in general, an increase in $\epsilon$ results in an increase in the positive difference between the estimated gate fidelity and the gate fidelity determined from process tomography results.

\subsection{Noisier gates}\label{sec:expnoisier}

We next investigate the extent to which our measurement-based interleaved randomised benchmarking protocol is able to detect noise variations in different measurement-based implementations of a gate.  To this end, we artificially increase noise in the measurement-based implementations of the Hadamard gate and the $T$ gate.  One option is to perform the single-qubit measurements in these measurement-based implementations at measurement angles which deviate from the required measurement angles.  However, this has the disadvantage that it is not easy to predict whether a measurement-based gate will be more or less noisy than the measurement-based 2-design used to estimate its fidelity.  We therefore artificially increase noise in the measurement-based implementations of the Hadamard gate and the $T$ gate by increasing the length of the linear cluster states used in the implementations.

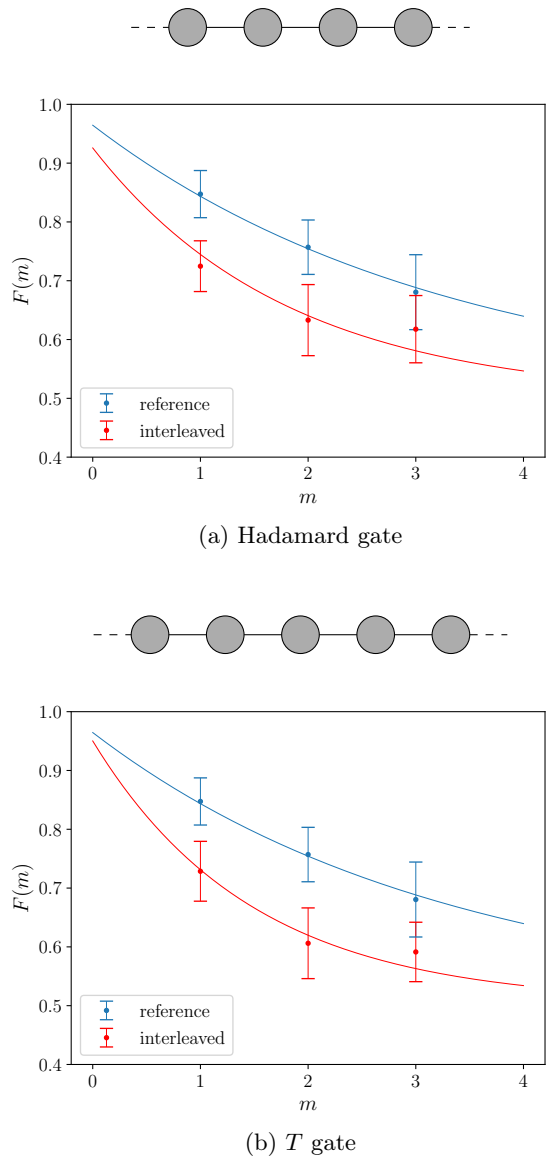
\begin{figure}
    \centering
    \begin{subfigure}[b]{.48\textwidth}
        \centering
        \begin{tikzpicture}
        \filldraw[fill=gray!65, draw=black] (0, 0) circle[radius=0.25];
        \filldraw[fill=gray!65, draw=black] (1, 0) circle[radius=0.25];
        \filldraw[fill=gray!65, draw=black] (2, 0) circle[radius=0.25];
        \filldraw[fill=gray!65, draw=black] (3, 0) circle[radius=0.25];
        \draw[dashed] (-0.75, 0) -- (-0.25, 0);
        \draw (0.25, 0) -- (0.75, 0);
        \draw (1.25, 0) -- (1.75, 0);
        \draw (2.25, 0) -- (2.75, 0);
        \draw[dashed] (3.25, 0) -- (3.75, 0);
        \end{tikzpicture}
        \includegraphics[scale=0.5]{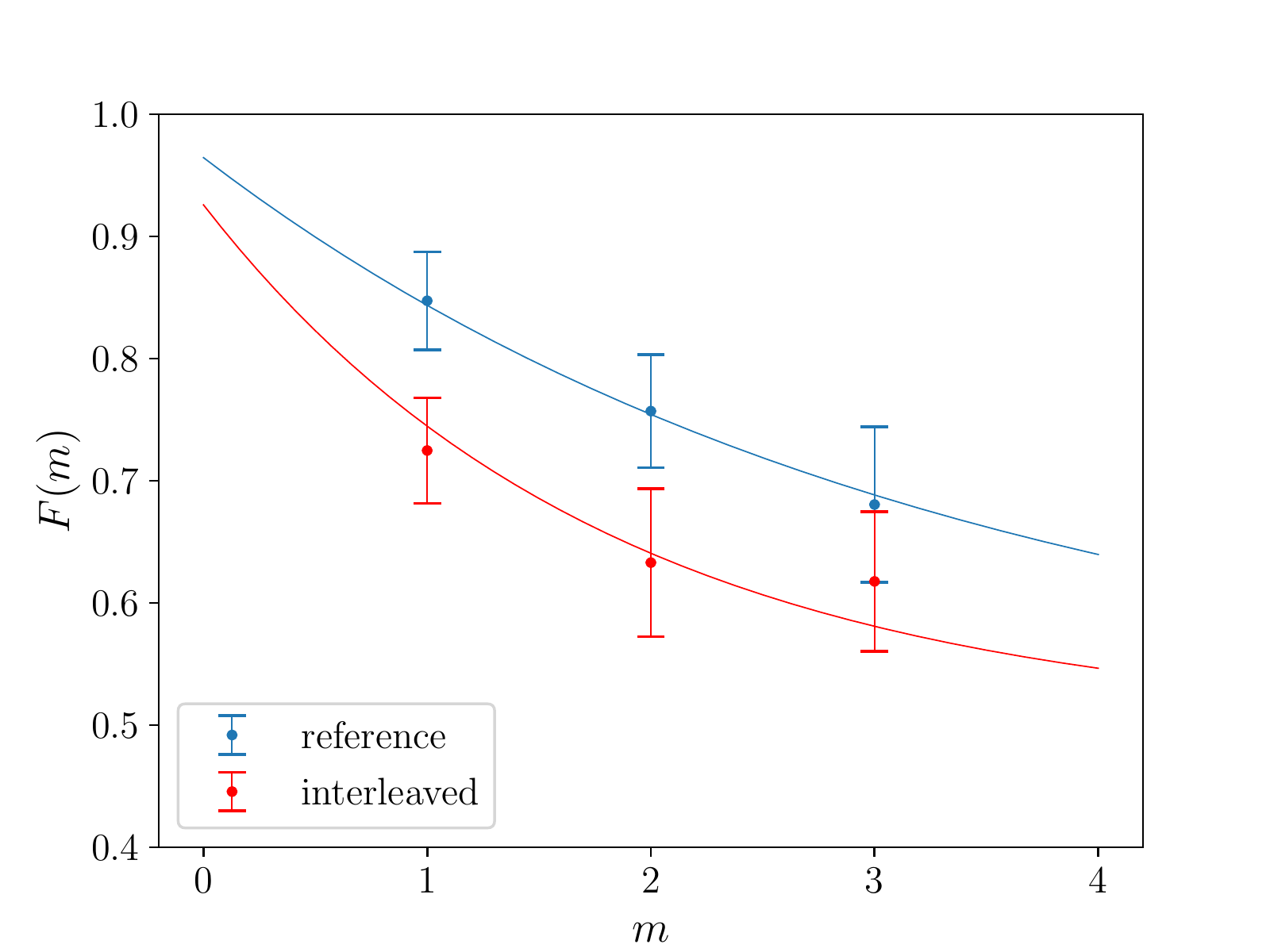}
        \caption{Hadamard gate}
    \end{subfigure}
    
    \vspace{0.75cm}
    
    \begin{subfigure}[b]{.48\textwidth}
        \centering
        \begin{tikzpicture}
        \filldraw[fill=gray!65, draw=black] (0, 0) circle[radius=0.25];
        \filldraw[fill=gray!65, draw=black] (1, 0) circle[radius=0.25];
        \filldraw[fill=gray!65, draw=black] (2, 0) circle[radius=0.25];
        \filldraw[fill=gray!65, draw=black] (3, 0) circle[radius=0.25];
        \filldraw[fill=gray!65, draw=black] (4, 0) circle[radius=0.25];
        \draw[dashed] (-0.75, 0) -- (-0.25, 0);
        \draw (0.25, 0) -- (0.75, 0);
        \draw (1.25, 0) -- (1.75, 0);
        \draw (2.25, 0) -- (2.75, 0);
        \draw (3.25, 0) -- (3.75, 0);
        \draw[dashed] (4.25, 0) -- (4.75, 0);
        \end{tikzpicture}
        \includegraphics[scale=0.5]{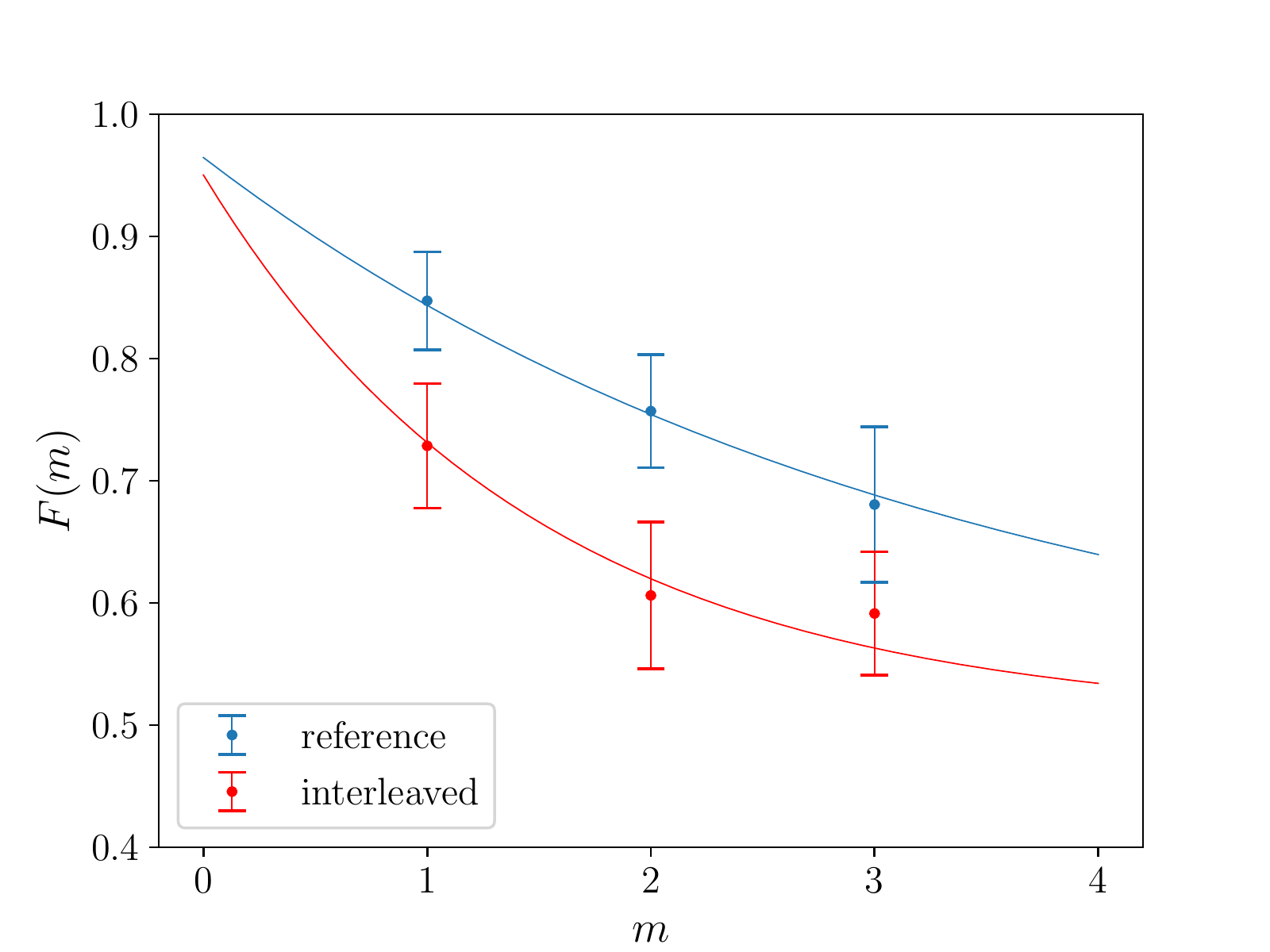}
        \caption{$T$ gate}
    \end{subfigure}
    \caption{Sequence fidelities $F(m)$ obtained for $m\in\{1,2,3\}$ to estimate the fidelity of the 4-qubit measurement-based implementation of the Hadamard gate and the 5-qubit measurement-based implementation of the $T$ gate, using the 5-qubit approximate measurement-based 2-design of Ref.~\cite{previous}, on the \textit{ibmq\_brooklyn} quantum processor.  Reference and interleaved sequence fidelities were fit to Eqs.~(\ref{eq:ref}) and (\ref{eq:int}) respectively (see \appendref{append:fitting}).  (a)~Hadamard gate shows the reference sequence fidelities (in blue) and the interleaved sequence fidelities for the 4-qubit Hadamard gate (in red).  (b)~$T$~gate shows the same reference sequence fidelities (in blue) and the interleaved sequence fidelities for the 5-qubit $T$ gate (in red).}
    \label{fig:noisemedium}
\end{figure}

Note that the 3-qubit linear cluster state with the measurement angles $\phi_1=0$ and $\phi_2=0$ implements the identity operation when all measurement outcomes are zero.  By appending this measurement-based implementation of the identity operation to the 2-qubit Hadamard gate and the 3-qubit $T$ gate given in \secref{sec:mbqcbackground}, we obtain measurement-based implementations of the Hadamard gate and the $T$ gate with longer cluster states.  In particular, when all measurement outcomes are zero, the 4-qubit linear cluster state with the measurement angles $\phi_1=0$, $\phi_2=0$ and $\phi_3=0$ implements the Hadamard gate and the 5-qubit linear cluster state with the measurement angles $\phi_1=\frac{\pi}{4}$, $\phi_2=0$, $\phi_3=0$ and $\phi_4=0$ implements the $T$ gate.  Furthermore, provided that all measurement outcomes are zero, the 6-qubit linear cluster state with the measurement angles $\phi_1=0$, $\phi_2=0$, $\phi_3=0$, $\phi_4=0$ and $\phi_5=0$ also implements the Hadamard gate and the 7-qubit linear cluster state with the measurement angles $\phi_1=\frac{\pi}{4}$, $\phi_2=0$, $\phi_3=0$, $\phi_4=0$, $\phi_5=0$ and $\phi_6=0$ also implements the $T$ gate.  When non-zero measurement outcomes do occur, the desired gate can be realised by simply applying the appropriate Pauli correction to the final qubit~\cite{MBQClc}, that is, these implementations all use fixed measurement angles and a Pauli correction and do not require adaptive measurement feedforward.

We estimated the fidelity of the 4-qubit and 6-qubit measurement-based implementations of the Hadamard gate and the 5-qubit and 7-qubit measurement-based implementations of the $T$ gate, using the 5-qubit approximate measurement-based 2-design of Ref.~\cite{previous}, on the \textit{ibmq\_brooklyn} quantum processor, a 65-qubit superconducting IBM quantum computer.  Details of the qubits used are provided in \appendref{append:qubitsnoisier}.  We once again implemented reference sequences with lengths $m\in\{1,2,3\}$ and interleaved sequences with $m\in\{1,2,3\}$, for each of the gates considered.  The \textit{ibmq\_brooklyn} quantum processor was used for these experiments, instead of the \textit{ibm\_hanoi} quantum processor, as the \textit{ibm\_hanoi} quantum processor has too few qubits to implement some of the required sequences, such as the interleaved sequence for the 7-qubit $T$ gate with $m=3$, which requires a 31-qubit linear cluster state, as shown in \figref{fig:brooklyntiny}.  Sequences were implemented, classical post-processing was done to determine the survival probability for each random sequence and sequence fidelities were calculated in the same way as in the experiments on the \textit{ibm\_hanoi} quantum processor (see \secref{sec:expimplementation}).

\begin{figure}[h!]
    \centering
    \begin{subfigure}[b]{.48\textwidth}
        \centering
        \begin{tikzpicture}
        \filldraw[fill=gray!65, draw=black] (0, 0) circle[radius=0.25];
        \filldraw[fill=gray!65, draw=black] (1, 0) circle[radius=0.25];
        \filldraw[fill=gray!65, draw=black] (2, 0) circle[radius=0.25];
        \filldraw[fill=gray!65, draw=black] (3, 0) circle[radius=0.25];
        \filldraw[fill=gray!65, draw=black] (4, 0) circle[radius=0.25];
        \filldraw[fill=gray!65, draw=black] (5, 0) circle[radius=0.25];
        \draw[dashed] (-0.75, 0) -- (-0.25, 0);
        \draw (0.25, 0) -- (0.75, 0);
        \draw (1.25, 0) -- (1.75, 0);
        \draw (2.25, 0) -- (2.75, 0);
        \draw (3.25, 0) -- (3.75, 0);
        \draw (4.25, 0) -- (4.75, 0);
        \draw[dashed] (5.25, 0) -- (5.75, 0);
        \end{tikzpicture}
        \includegraphics[scale=0.5]{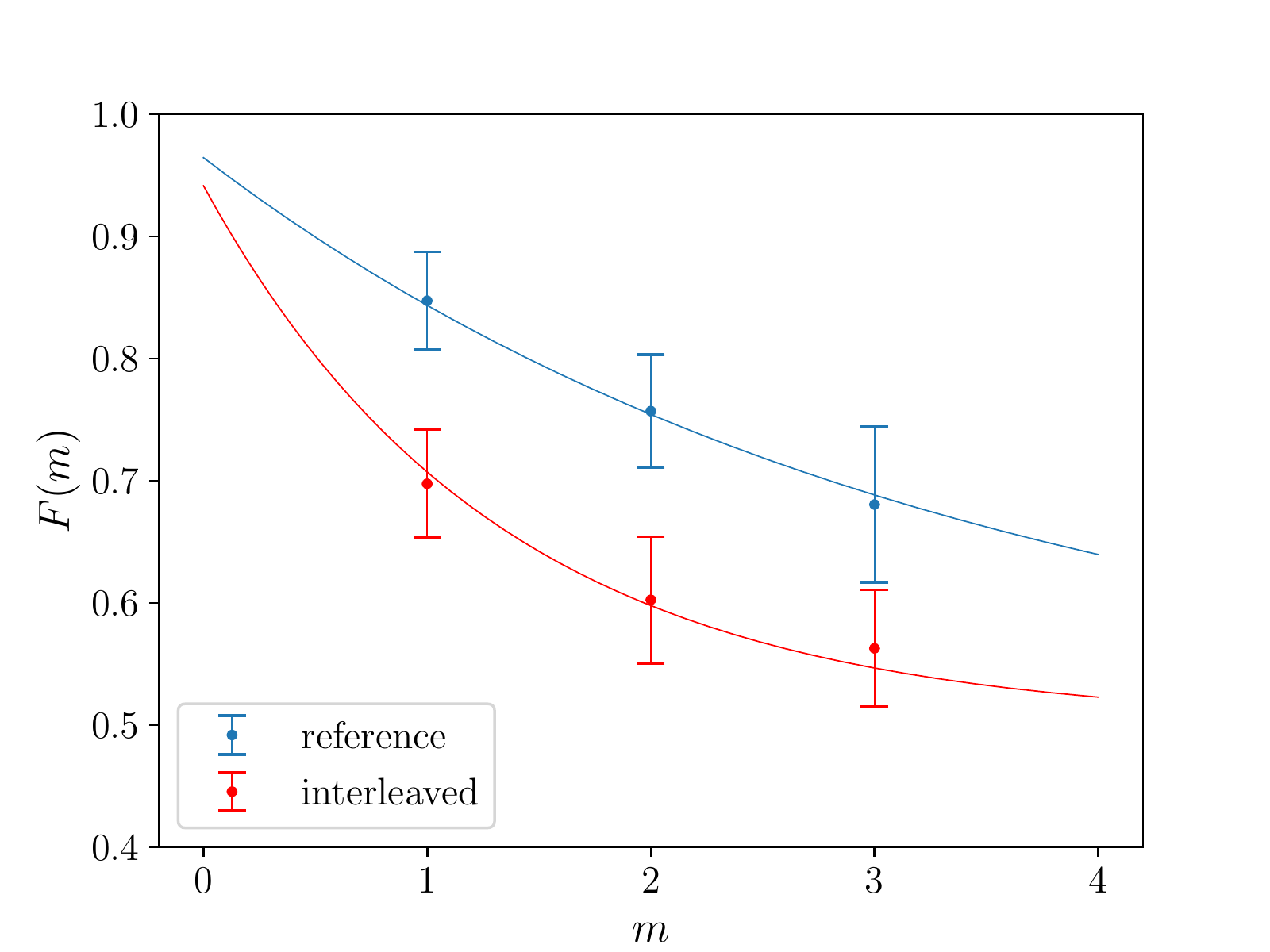}
        \caption{Hadamard gate}
    \end{subfigure}
    
    \vspace{0.75cm}
    
    \begin{subfigure}[b]{.48\textwidth}
        \centering
        \begin{tikzpicture}
        \filldraw[fill=gray!65, draw=black] (0, 0) circle[radius=0.25];
        \filldraw[fill=gray!65, draw=black] (1, 0) circle[radius=0.25];
        \filldraw[fill=gray!65, draw=black] (2, 0) circle[radius=0.25];
        \filldraw[fill=gray!65, draw=black] (3, 0) circle[radius=0.25];
        \filldraw[fill=gray!65, draw=black] (4, 0) circle[radius=0.25];
        \filldraw[fill=gray!65, draw=black] (5, 0) circle[radius=0.25];
        \filldraw[fill=gray!65, draw=black] (6, 0) circle[radius=0.25];
        \draw[dashed] (-0.75, 0) -- (-0.25, 0);
        \draw (0.25, 0) -- (0.75, 0);
        \draw (1.25, 0) -- (1.75, 0);
        \draw (2.25, 0) -- (2.75, 0);
        \draw (3.25, 0) -- (3.75, 0);
        \draw (4.25, 0) -- (4.75, 0);
        \draw (5.25, 0) -- (5.75, 0);
        \draw[dashed] (6.25, 0) -- (6.75, 0);
        \end{tikzpicture}
        \includegraphics[scale=0.5]{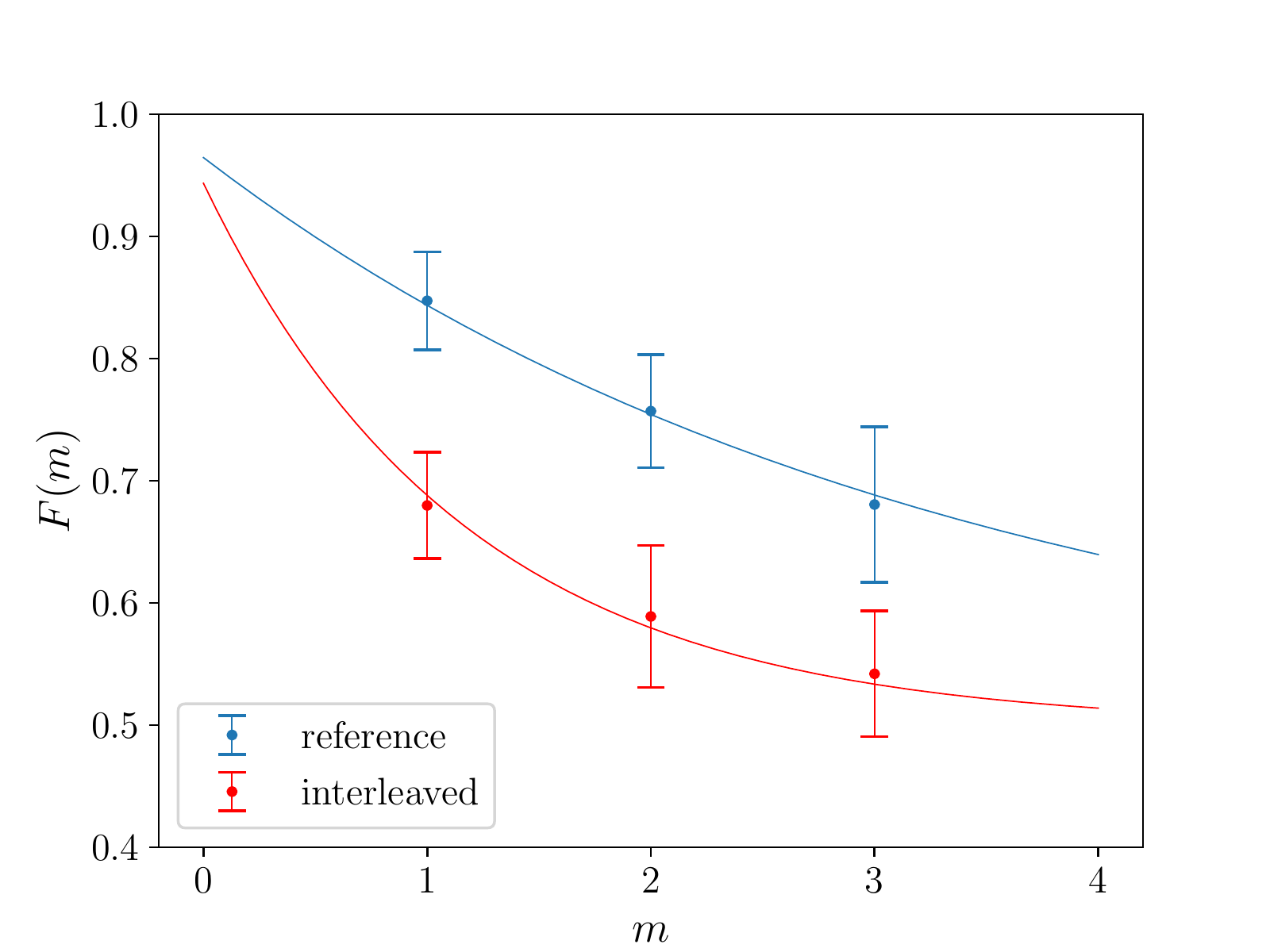}
        \caption{$T$ gate}
    \end{subfigure}
    \caption{Sequence fidelities $F(m)$ obtained for $m\in\{1,2,3\}$ to estimate the fidelity of the 6-qubit measurement-based implementation of the Hadamard gate and the 7-qubit measurement-based implementation of the $T$ gate, using the 5-qubit approximate measurement-based 2-design of Ref.~\cite{previous}, on the \textit{ibmq\_brooklyn} quantum processor.  Reference and interleaved sequence fidelities were fit to Eqs.~(\ref{eq:ref}) and (\ref{eq:int}) respectively (see \appendref{append:fitting}).  (a)~Hadamard gate shows the reference sequence fidelities from \figref{fig:noisemedium} (in blue) and the interleaved sequence fidelities for the 6-qubit Hadamard gate (in red).  (b)~$T$~gate shows the same reference sequence fidelities from \figref{fig:noisemedium} (in blue) and the interleaved sequence fidelities for the 7-qubit $T$ gate (in red).}
    \label{fig:noisehigh}
\end{figure}

The only difference in the experiments on the \textit{ibmq\_brooklyn} quantum processor is that for the interleaved sequences for the measurement-based gates defined here, we need not perform quantum state tomography for each of the possible measurement outcomes to infer the output state for each random interleaved sequence.  This can be understood as follows.  For the 2-qubit Hadamard gate and the 3-qubit $T$ gate there is a one-to-one correspondence between the outcomes of single-qubit measurements in the measurement-based implementations and the random byproducts that result from omitting the Pauli corrections.  This results in a one-to-one correspondence between random measurement outcomes and random interleaved sequences, since the random interleaved sequences consist of random unitary operators interleaved with random byproducts.  In contrast, there are 8, 16, 32 and 64 possible outcomes for the single-qubit measurements performed in the respective 4-qubit, 5-qubit, 6-qubit and 7-qubit measurement-based implementations defined here, but there are only four possible byproducts, corresponding to the four possible Pauli corrections, for each implementation.  As a result, there is no longer a one-to-one correspondence between random measurement outcomes and random interleaved sequences.  Therefore, to infer the output state for each random interleaved sequence, we need not perform quantum state tomography for each of the possible measurement outcomes, since tomography data obtained for different measurement outcomes which correspond to the same random interleaved sequence can be grouped.

We note that, for the interleaved sequences for the measurement-based gates defined here, it is in fact not feasible to perform quantum state tomography for each of the possible measurement outcomes, since the number of possible measurement outcomes grows exponentially with the length of the cluster state.  For example, the circuit implementing the interleaved sequence for the 7-qubit $T$ gate with $m=3$ would need to be run $3(2^{30})/16\approx201$ million times to obtain the data required to perform tomography for each of the possible measurement outcomes.  By performing tomography for each random interleaved sequence, and grouping tomography data obtained for different measurement outcomes corresponding to the same random sequence, we were able to drastically reduce the number of times each circuit had to be run.  In particular, to obtain the data required to determine the interleaved sequence fidelity for a given $m$, for the 4-qubit, 5-qubit, 6-qubit and 7-qubit measurement-based gates defined here, the relevant circuit was run the same number of times as for the 3-qubit $T$ gate (see \secref{sec:expimplementation}), since the number of possible byproducts, and therefore the number of random interleaved sequences, is the same as for the 3-qubit $T$ gate.

Sequence fidelities obtained to estimate the fidelity of the 4-qubit Hadamard gate and the 5-qubit $T$ gate, using the 5-qubit approximate measurement-based 2-design of Ref.~\cite{previous}, on the \textit{ibmq\_brooklyn} quantum processor are shown in \figref{fig:noisemedium}.  Furthermore, sequence fidelities obtained to estimate the fidelity of the 6-qubit Hadamard gate and the 7-qubit $T$ gate are shown in \figref{fig:noisehigh}.  \tabref{tab:fidelities} shows the estimated gate fidelities, as well as the range of Haar-averaged gate fidelities calculated from process tomography results (see \appendref{append:channeltomo}).  Both the estimated gate fidelities and the gate fidelities obtained from process tomography results decrease as the length of the cluster state in the implementation increases, which reflects the expected increase in noise resulting from increasing the length of the cluster state.  This shows that, even with very little data, our measurement-based interleaved randomised benchmarking protocol is able to detect large noise variations in different measurement-based implementations of a gate.

\begin{table}[]
\begin{tabular}{|l|l|l|}
\hline
\textbf{Gate}         & \textbf{Est Fidelity} & \textbf{Fidelity Range} \\ \hline
4-qubit Hadamard gate & 0.894$\pm$0.092             & 0.831--0.895            \\ \hline
5-qubit $T$ gate      & 0.849$\pm$0.085             & 0.821--0.851            \\ \hline
6-qubit Hadamard gate & 0.820$\pm$0.079             & 0.693--0.835            \\ \hline
7-qubit $T$ gate      & 0.791$\pm$0.081             & 0.702--0.803            \\ \hline
\end{tabular}
\caption{Haar-averaged gate fidelities obtained for the different measurement-based gates on the \textit{ibmq\_brooklyn} quantum processor.  `Est Fidelity' shows the fidelity estimated using our measurement-based interleaved randomised benchmarking protocol.  `Fidelity Range' shows the range of fidelities calculated from process tomography results obtained for the three different sets of qubits used in the interleaved sequences (see \appendref{append:channeltomo}).}
\label{tab:fidelities}
\end{table}

\section{Conclusion}\label{sec:conclusion} 

We proposed an interleaved randomised benchmarking protocol for measurement-based quantum computers, which is an extension of the standard randomised benchmarking protocol proposed for measurement-based quantum computers by Alexander \textit{et al.}~\cite{MBSRB}.  In our measurement-based interleaved randomised benchmarking protocol, any single-qubit measurement-based 2-design can be used to estimate the fidelity of any single-qubit measurement-based gate.  Future work could involve developing interleaved randomised benchmarking protocols in which multi-qubit measurement-based 2-designs~\cite{MBT2} can be used to estimate the fidelity of multi-qubit measurement-based gates.  Obstacles that would need to be overcome in this regard include computing the inverse of a random multi-qubit sequence, since multi-qubit systems cannot generally be efficiently simulated on a classical computer, and applying this inverse in such a way that noise from the inverse is independent of the sequence, since the inverse of a multi-qubit sequence cannot be applied by performing a single-qubit measurement basis rotation.  Random measurement-based Clifford gates may help in this regard, where they may be used to bound the fidelity of non-Clifford gates that form a universal set~\cite{IRBgeneral}.  Another option is to combine Clifford group and Pauli group gates within the interleaved sequence of a measurement-based non-Clifford gate~\cite{IRB5}.  Future work investigating the implications of performing randomised benchmarking with an approximate 2-design, as opposed to an exact 2-design, is also needed, since it is still unknown whether exact multi-qubit measurement-based 2-designs exist.

We tested our protocol by using the approximate measurement-based 2-design proposed in our previous work~\cite{previous} to estimate the fidelity of measurement-based implementations of the Hadamard gate and the $T$ gate (a universal single-qubit set) on the remotely accessible IBM superconducting quantum computers.  Since IBM processors did not support dynamic circuit execution at the time of performing the experiments, our protocol had to be adjusted to enable implementation on the superconducting quantum hardware available at the time.  In particular, it was not possible to implement the inverse of a random sequence as a quantum mechanical operation without dynamic circuit execution, since the sequence, and therefore its inverse, is only known after the required single-qubit measurements have been performed.  By preparing linear cluster states of up to 31 qubits, performing single-qubit measurements on all but the final qubit, performing quantum state tomography on the final qubit to infer the output state for each random sequence, and by using classical post-processing to apply the inverse of each sequence and extract the survival probability, we were able to determine reference sequence fidelities for sequence lengths $m\in\{1,2,3\}$ and interleaved sequence fidelities for $m\in\{1,2,3\}$ for each of the gates considered.

In our adjusted protocol, the resources required to obtain the data needed to determine the sequence fidelity scale exponentially with the length of the sequence, which is why the number of sequence fidelities determined here is much smaller than in typical interleaved randomised benchmarking experiments~\cite{IRB1, RBsuperconducting3}.  Hence, even though our adjustments have enabled us to obtain data for a proof-of-concept demonstration, dynamic circuit execution remains essential for the efficient scaling of our measurement-based interleaved randomised benchmarking protocol.  We note that our measurement-based interleaved randomised benchmarking protocol could be implemented as presented in \secref{sec:protocol}, without any adjustments, on a measurement-based architecture or a circuit-based architecture which supports dynamic circuit execution.  A measurement-based architecture or a circuit-based architecture which supports dynamic circuit execution would therefore eliminate the exponential scaling with sequence length as well as the need for quantum state tomography and classical post-processing.  Recent work on circumventing dynamic circuit execution on IBM processors for the measurement-based model via a delayed choice strategy~\cite{MBQCIBM} is an interesting direction, although such a strategy comes at the expense of adding further entangling gates, which may introduce additional noise.  Dynamic circuit execution is thus an important addition to IBM processors for the measurement-based model, and the recent addition thereof opens up many opportunities for quantum computing in the measurement-based model.

In all the experiments, estimated gate fidelities show good agreement with gate fidelities calculated from process tomography results.  Even though some gate fidelities are slightly over-estimated and the uncertainties of estimated gate fidelities are larger than the uncertainties of those calculated from process tomography results, the experimental results clearly demonstrate the robustness of interleaved randomised benchmarking if one takes into consideration that we used a very weak approximate 2-design~\cite{previous}, the implementation of this 2-design on the IBM quantum processors did not even pass our test for an approximate 2-design for all states~\cite{previous}, noise from this 2-design is not entirely gate-independent~\cite{previous} and the fitting parameters were inferred from very few sequence fidelities.  Furthermore, by artificially increasing noise in the measurement-based implementations of the Hadamard gate and the $T$ gate, we were able to show that, even with very little data, our measurement-based interleaved randomised benchmarking protocol is able to detect large noise variations in different measurement-based implementations of a gate.  Our work highlights the usefulness of cloud-based superconducting systems for single-qubit measurement-based quantum computing and shows how to practically characterise noisy quantum logic gates in this setting.  In future experiments, our protocol could be implemented on custom-built linear optical systems~\cite{MBQCexam1, MBQCexam2, MBQCexam3, photonic1, photonic2, photonic3, photonic4}, one of the most promising physical systems for measurement-based quantum computing.

\begin{acknowledgements}
We acknowledge the use of IBM Quantum services for this work. The views expressed are those of the authors, and do not reflect the official policy or position of IBM or the IBM Quantum team.  We thank Taariq Surtee and Barry Dwolatzky at the University of Witwatersrand and Ismail Akhalwaya at IBM Research Africa for access to the IBM processors through the Q Network and Africa Research Universities Alliance.  This research was supported by the South African National Research Foundation, the University of Stellenbosch, and the South African Research Chair Initiative of the Department of Science and Technology and National Research Foundation.
\end{acknowledgements}


\appendix

\onecolumngrid

\newpage

\section{Qubits used for fidelity estimation of universal gates}\label{append:qubitsuniversal} 

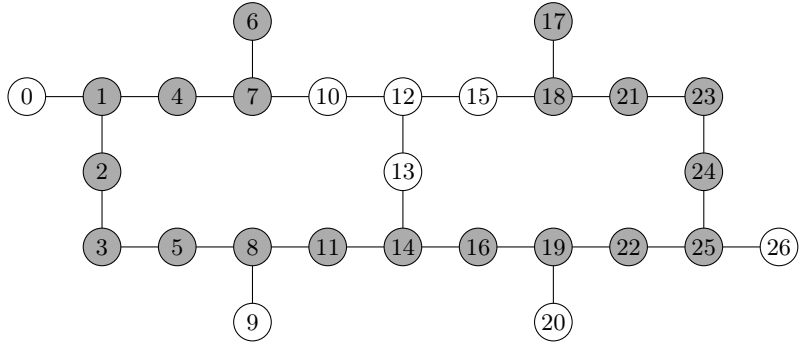
\begin{figure*}[h]
    \centering
    \begin{tikzpicture}
    \filldraw[fill=gray!65, draw=black] (3, 1) circle[radius=0.25] node {6};
    \filldraw[fill=gray!65, draw=black] (7, 1) circle[radius=0.25] node {17};
    \draw (0, 0) circle[radius=0.25] node {0};
    \filldraw[fill=gray!65, draw=black] (1, 0) circle[radius=0.25] node {1};
    \filldraw[fill=gray!65, draw=black] (2, 0) circle[radius=0.25] node {4};
    \filldraw[fill=gray!65, draw=black] (3, 0) circle[radius=0.25] node {7};
    \draw (4, 0) circle[radius=0.25] node {10};
    \draw (5, 0) circle[radius=0.25] node {12};
    \draw (6, 0) circle[radius=0.25] node {15};
    \filldraw[fill=gray!65, draw=black] (7, 0) circle[radius=0.25] node {18};
    \filldraw[fill=gray!65, draw=black] (8, 0) circle[radius=0.25] node {21};
    \filldraw[fill=gray!65, draw=black] (9, 0) circle[radius=0.25] node {23};
    \filldraw[fill=gray!65, draw=black] (1, -1) circle[radius=0.25] node {2};
    \draw (5, -1) circle[radius=0.25] node {13};
    \filldraw[fill=gray!65, draw=black] (9, -1) circle[radius=0.25] node {24};
    \filldraw[fill=gray!65, draw=black] (1, -2) circle[radius=0.25] node {3};
    \filldraw[fill=gray!65, draw=black] (2, -2) circle[radius=0.25] node {5};
    \filldraw[fill=gray!65, draw=black] (3, -2) circle[radius=0.25] node {8};
    \filldraw[fill=gray!65, draw=black] (4, -2) circle[radius=0.25] node {11};
    \filldraw[fill=gray!65, draw=black] (5, -2) circle[radius=0.25] node {14};
    \filldraw[fill=gray!65, draw=black] (6, -2) circle[radius=0.25] node {16};
    \filldraw[fill=gray!65, draw=black] (7, -2) circle[radius=0.25] node {19};
    \filldraw[fill=gray!65, draw=black] (8, -2) circle[radius=0.25] node {22};
    \filldraw[fill=gray!65, draw=black] (9, -2) circle[radius=0.25] node {25};
    \draw (10, -2) circle[radius=0.25] node {26};
    \draw (3, -3) circle[radius=0.25] node {9};
    \draw (7, -3) circle[radius=0.25] node {20};
    \draw (3, 0.25) -- (3, 0.75);
    \draw (7, 0.25) -- (7, 0.75);
    \draw (0.25, 0) -- (0.75, 0);
    \draw (1.25, 0) -- (1.75, 0);
    \draw (2.25, 0) -- (2.75, 0);
    \draw (3.25, 0) -- (3.75, 0);
    \draw (4.25, 0) -- (4.75, 0);
    \draw (5.25, 0) -- (5.75, 0);
    \draw (6.25, 0) -- (6.75, 0);
    \draw (7.25, 0) -- (7.75, 0);
    \draw (8.25, 0) -- (8.75, 0);
    \draw (1, -0.25) -- (1, -0.75);
    \draw (5, -0.25) -- (5, -0.75);
    \draw (9, -0.25) -- (9, -0.75);
    \draw (1, -1.25) -- (1, -1.75);
    \draw (5, -1.25) -- (5, -1.75);
    \draw (9, -1.25) -- (9, -1.75);
    \draw (1.25, -2) -- (1.75, -2);
    \draw (2.25, -2) -- (2.75, -2);
    \draw (3.25, -2) -- (3.75, -2);
    \draw (4.25, -2) -- (4.75, -2);
    \draw (5.25, -2) -- (5.75, -2);
    \draw (6.25, -2) -- (6.75, -2);
    \draw (7.25, -2) -- (7.75, -2);
    \draw (8.25, -2) -- (8.75, -2);
    \draw (9.25, -2) -- (9.75, -2);
    \draw (3, -2.25) -- (3, -2.75);
    \draw (7, -2.25) -- (7, -2.75);
    \end{tikzpicture}
    \caption{Qubit topology of the \textit{ibm\_hanoi} quantum processor.  The connecting lines between qubits indicate the qubit pairs for which the $CX$ gate is supported at the hardware level.  The qubits used in the implementations are shaded grey.}
    \label{fig:hanoi}
\end{figure*}

To perform fidelity estimation for a universal single-qubit set, we implemented reference sequences with lengths $m\in\{1,2,3\}$ and interleaved sequences with $m\in\{1,2,3\}$, for both the 2-qubit Hadamard gate and the 3-qubit $T$ gate, on the \textit{ibm\_hanoi} quantum processor.  The qubit topology of the \textit{ibm\_hanoi} quantum processor is shown in \figref{fig:hanoi}, with the qubits used in the implementations shaded grey.  These qubits were chosen for their low error rates compared to other qubits.  For each $m$ (reference or interleaved), the required linear cluster state was prepared along the length of shaded qubits, starting at qubit 17.  For example, to implement the interleaved sequence for the 3-qubit $T$ gate with $m=1$, the required 7-qubit linear cluster state was prepared on the qubits 17, 18, 21, 23, 24, 25 and 22, single-qubit measurements were performed on the qubits 17, 18, 21, 23, 24 and 25, and quantum state tomography was performed on qubit 22.  All nineteen shaded qubits, from qubit 17 through to qubit 6, were used to implement the interleaved sequence for the 3-qubit $T$ gate with $m=3$, which required a 19-qubit linear cluster state.

The required data was obtained over a period of 16 days on the \textit{ibm\_hanoi} quantum processor.  On each day on which circuits were run, the relevant calibration information was recorded at 19:00 GMT.  \tabref{tab:calibrationhanoi} shows the average of the calibration information recorded during this period, with the uncertainties given by the standard deviation.

\begin{table}[h]
\centering
\begin{minipage}{11.5cm}
\centering
\begin{tabular}{|l|l|l|l|l|}
\hline
\textbf{Qubit} & $\boldsymbol{T_1}\ \boldsymbol{(}\boldsymbol{\mu}\mathbf{s}\boldsymbol{)}$ & $\boldsymbol{T_2}\ \boldsymbol{(}\boldsymbol{\mu}\mathbf{s}\boldsymbol{)}$ & $\boldsymbol{\sqrt{X}}$ \textbf{Error} & \textbf{Readout Error} \\ \hline
17             & 135.10$\pm$25.32 & $\,\,\,$75.47$\pm$04.72  & 0.000316$\pm$0.000045     & 0.0149$\pm$0.0039      \\ \hline
18             & 174.96$\pm$33.66 & 173.48$\pm$38.68 & 0.000360$\pm$0.000055     & 0.0139$\pm$0.0053      \\ \hline
21             & 137.16$\pm$20.95 & $\,\,\,$24.48$\pm$01.93  & 0.000268$\pm$0.000023     & 0.0124$\pm$0.0027      \\ \hline
23             & 152.68$\pm$33.73 & $\,\,\,$45.98$\pm$02.76  & 0.000363$\pm$0.000052     & 0.0223$\pm$0.0079      \\ \hline
24             & 153.91$\pm$27.54 & $\,\,\,$31.14$\pm$04.12  & 0.000355$\pm$0.000048     & 0.0129$\pm$0.0010      \\ \hline
25             & 178.63$\pm$30.70 & $\,\,\,$70.14$\pm$13.07  & 0.000148$\pm$0.000040     & 0.0171$\pm$0.0028      \\ \hline
22             & 162.43$\pm$40.46 & 102.68$\pm$24.47 & 0.000212$\pm$0.000128     & 0.0134$\pm$0.0051      \\ \hline
19             & 153.41$\pm$40.11 & 149.65$\pm$56.73 & 0.000133$\pm$0.000016     & 0.0065$\pm$0.0007      \\ \hline
16             & 114.35$\pm$45.09 & 143.93$\pm$59.36 & 0.000238$\pm$0.000117     & 0.0142$\pm$0.0012      \\ \hline
14             & 148.31$\pm$34.02 & $\,\,\,$27.22$\pm$00.00  & 0.000456$\pm$0.000522     & 0.0113$\pm$0.0053      \\ \hline
11             & 158.01$\pm$40.07 & 197.11$\pm$47.48 & 0.000129$\pm$0.000027     & 0.0145$\pm$0.0012      \\ \hline
8              & 170.04$\pm$38.89 & 260.68$\pm$69.31 & 0.000145$\pm$0.000065     & 0.0117$\pm$0.0020      \\ \hline
5              & 136.51$\pm$24.61 & 148.65$\pm$39.80 & 0.000550$\pm$0.000221     & 0.0127$\pm$0.0053      \\ \hline
3              & 164.39$\pm$38.07 & 247.30$\pm$72.86 & 0.000120$\pm$0.000016     & 0.0070$\pm$0.0007      \\ \hline
2              & 149.60$\pm$32.88 & 236.10$\pm$42.03 & 0.000120$\pm$0.000025     & 0.0071$\pm$0.0033      \\ \hline
1              & 165.78$\pm$31.71 & 149.88$\pm$43.46 & 0.000219$\pm$0.000071     & 0.0115$\pm$0.0059      \\ \hline
4              & 133.92$\pm$42.34 & $\,\,\,$15.79$\pm$00.00  & 0.000167$\pm$0.000018     & 0.0080$\pm$0.0011      \\ \hline
7              & 175.04$\pm$36.21 & 216.79$\pm$46.00 & 0.000129$\pm$0.000014     & 0.0138$\pm$0.0020      \\ \hline
6              & 148.74$\pm$30.86 & 184.66$\pm$67.86 & 0.000336$\pm$0.000223     & 0.0155$\pm$0.0071      \\ \hline
\end{tabular}
\end{minipage}%
\begin{minipage}{5.5cm}
\centering
\begin{tabular}{|l|l|}
\hline
\textbf{Qubit Pair} & $\boldsymbol{C}\boldsymbol{X}$ \textbf{Error} \\ \hline
17--18              & 0.00691$\pm$0.00093 \\ \hline
18--21              & 0.00545$\pm$0.00100 \\ \hline
21--23              & 0.01434$\pm$0.00174 \\ \hline
23--24              & 0.02561$\pm$0.00231 \\ \hline
24--25              & 0.01961$\pm$0.00434 \\ \hline
25--22              & 0.00696$\pm$0.00170 \\ \hline
22--19              & 0.00919$\pm$0.00163 \\ \hline
19--16              & 0.00713$\pm$0.00270 \\ \hline
16--14              & 0.01480$\pm$0.00534 \\ \hline
14--11              & 0.01167$\pm$0.01659 \\ \hline
11--8               & 0.00397$\pm$0.00088 \\ \hline
8--5                & 0.01939$\pm$0.00614 \\ \hline
5--3                & 0.00622$\pm$0.00199 \\ \hline
3--2                & 0.00646$\pm$0.00088 \\ \hline
2--1                & 0.00348$\pm$0.00040 \\ \hline
1--4                & 0.00709$\pm$0.00129 \\ \hline
4--7                & 0.01015$\pm$0.00119 \\ \hline
7--6                & 0.00541$\pm$0.00090 \\ \hline
\end{tabular}
\end{minipage}
\caption{Calibration information for the \textit{ibm\_hanoi} quantum processor averaged over the time period during which circuits were run and data was obtained.  The single-qubit calibration information for the relevant qubits is shown on the left.  $T_1$ and $T_2$ are the relaxation and dephasing times respectively of the qubits.  The $CX$ error rates for relevant qubit pairs are shown on the right.}
\label{tab:calibrationhanoi}
\end{table}

\section{Fitting procedure}\label{append:fitting} 

A Monte Carlo method, which takes uncertainties into account, was used to fit the reference and interleaved sequence fidelities to the exponential decay model given by Eqs.~(\ref{eq:ref}) and (\ref{eq:int}), respectively.  For each sequence fidelity, one million points were sampled from a Gaussian distribution with mean equal to the sequence fidelity and standard deviation equal to the uncertainty in the sequence fidelity.  For a given reference or interleaved sequence, sampled points obtained for the sequence fidelities for $m\in\{1,2,3\}$ were fit to \equatref{eq:ref} or \equatref{eq:int}, to obtain one million values for each fitting parameter.  A given fitting parameter, inferred for a given reference or interleaved sequence, is the average of these one million values, and the uncertainty is given by the standard deviation.

Since each set of fitting parameters ($p$, $A$ and $B$) was inferred from only three sequence fidelities, tight fitting constraints were imposed to ensure that the fitting parameters inferred from the data are physical.  In particular, the constraint $B\in [0.48, 0.52]$ was imposed to ensure that the asymptote of the exponential decay curve is close to $0.5$, as we expect it to be for depolarising noise.  Furthermore, the constraint $A\in [0.4, 0.5]$ was imposed to ensure that the $y$-intercept of the exponential decay curve (which is given by $A+B$) is slightly less than one, as we expect it to be when state preparation and measurement errors are present, but do not dominate.

Finally, a similar Monte Carlo method was used to estimate the Haar-averaged gate fidelity of a given measurement-based gate from the appropriate reference depolarising noise parameter $p_{\text{ref}}$ and interleaved depolarising noise parameter $p_{\text{int}}$.  One million points were sampled from a Gaussian distribution with mean equal to $p_{\text{ref}}$ and standard deviation equal to the uncertainty in $p_{\text{ref}}$, and similarly for $p_{\text{int}}$.  One million values were then obtained for the estimated gate fidelity, from the sampled points for $p_{\text{ref}}$ and $p_{\text{int}}$, using \equatref{eq:fidelity}.  The estimated gate fidelity for a given measurement-based gate is the average of these one million values, and the uncertainty is given by the standard deviation.

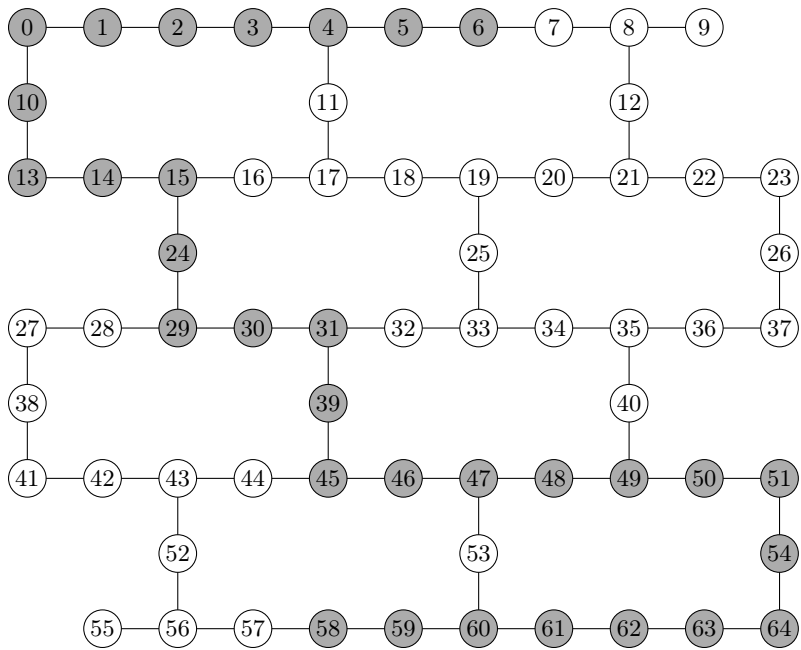
\begin{figure*}
    \centering
    \begin{tikzpicture}
    \filldraw[fill=gray!65, draw=black] (0, 0) circle[radius=0.25] node {0};
    \filldraw[fill=gray!65, draw=black] (1, 0) circle[radius=0.25] node {1};
    \filldraw[fill=gray!65, draw=black] (2, 0) circle[radius=0.25] node {2};
    \filldraw[fill=gray!65, draw=black] (3, 0) circle[radius=0.25] node {3};
    \filldraw[fill=gray!65, draw=black] (4, 0) circle[radius=0.25] node {4};
    \filldraw[fill=gray!65, draw=black] (5, 0) circle[radius=0.25] node {5};
    \filldraw[fill=gray!65, draw=black] (6, 0) circle[radius=0.25] node {6};
    \draw (7, 0) circle[radius=0.25] node {7};
    \draw (8, 0) circle[radius=0.25] node {8};
    \draw (9, 0) circle[radius=0.25] node {9};
    \filldraw[fill=gray!65, draw=black] (0, -1) circle[radius=0.25] node {10};
    \draw (4, -1) circle[radius=0.25] node {11};
    \draw (8, -1) circle[radius=0.25] node {12};
    \filldraw[fill=gray!65, draw=black] (0, -2) circle[radius=0.25] node {13};
    \filldraw[fill=gray!65, draw=black] (1, -2) circle[radius=0.25] node {14};
    \filldraw[fill=gray!65, draw=black] (2, -2) circle[radius=0.25] node {15};
    \draw (3, -2) circle[radius=0.25] node {16};
    \draw (4, -2) circle[radius=0.25] node {17};
    \draw (5, -2) circle[radius=0.25] node {18};
    \draw (6, -2) circle[radius=0.25] node {19};
    \draw (7, -2) circle[radius=0.25] node {20};
    \draw (8, -2) circle[radius=0.25] node {21};
    \draw (9, -2) circle[radius=0.25] node {22};
    \draw (10, -2) circle[radius=0.25] node {23};
    \filldraw[fill=gray!65, draw=black] (2, -3) circle[radius=0.25] node {24};
    \draw (6, -3) circle[radius=0.25] node {25};
    \draw (10, -3) circle[radius=0.25] node {26};
    \draw (0, -4) circle[radius=0.25] node {27};
    \draw (1, -4) circle[radius=0.25] node {28};
    \filldraw[fill=gray!65, draw=black] (2, -4) circle[radius=0.25] node {29};
    \filldraw[fill=gray!65, draw=black] (3, -4) circle[radius=0.25] node {30};
    \filldraw[fill=gray!65, draw=black] (4, -4) circle[radius=0.25] node {31};
    \draw (5, -4) circle[radius=0.25] node {32};
    \draw (6, -4) circle[radius=0.25] node {33};
    \draw (7, -4) circle[radius=0.25] node {34};
    \draw (8, -4) circle[radius=0.25] node {35};
    \draw (9, -4) circle[radius=0.25] node {36};
    \draw (10, -4) circle[radius=0.25] node {37};
    \draw (0, -5) circle[radius=0.25] node {38};
    \filldraw[fill=gray!65, draw=black] (4, -5) circle[radius=0.25] node {39};
    \draw (8, -5) circle[radius=0.25] node {40};
    \draw (0, -6) circle[radius=0.25] node {41};
    \draw (1, -6) circle[radius=0.25] node {42};
    \draw (2, -6) circle[radius=0.25] node {43};
    \draw (3, -6) circle[radius=0.25] node {44};
    \filldraw[fill=gray!65, draw=black] (4, -6) circle[radius=0.25] node {45};
    \filldraw[fill=gray!65, draw=black] (5, -6) circle[radius=0.25] node {46};
    \filldraw[fill=gray!65, draw=black] (6, -6) circle[radius=0.25] node {47};
    \filldraw[fill=gray!65, draw=black] (7, -6) circle[radius=0.25] node {48};
    \filldraw[fill=gray!65, draw=black] (8, -6) circle[radius=0.25] node {49};
    \filldraw[fill=gray!65, draw=black] (9, -6) circle[radius=0.25] node {50};
    \filldraw[fill=gray!65, draw=black] (10, -6) circle[radius=0.25] node {51};
    \draw (2, -7) circle[radius=0.25] node {52};
    \draw (6, -7) circle[radius=0.25] node {53};
    \filldraw[fill=gray!65, draw=black] (10, -7) circle[radius=0.25] node {54};
    \draw (1, -8) circle[radius=0.25] node {55};
    \draw (2, -8) circle[radius=0.25] node {56};
    \draw (3, -8) circle[radius=0.25] node {57};
    \filldraw[fill=gray!65, draw=black] (4, -8) circle[radius=0.25] node {58};
    \filldraw[fill=gray!65, draw=black] (5, -8) circle[radius=0.25] node {59};
    \filldraw[fill=gray!65, draw=black] (6, -8) circle[radius=0.25] node {60};
    \filldraw[fill=gray!65, draw=black] (7, -8) circle[radius=0.25] node {61};
    \filldraw[fill=gray!65, draw=black] (8, -8) circle[radius=0.25] node {62};
    \filldraw[fill=gray!65, draw=black] (9, -8) circle[radius=0.25] node {63};
    \filldraw[fill=gray!65, draw=black] (10, -8) circle[radius=0.25] node {64};
    \draw (0.25, 0) -- (0.75, 0);
    \draw (1.25, 0) -- (1.75, 0);
    \draw (2.25, 0) -- (2.75, 0);
    \draw (3.25, 0) -- (3.75, 0);
    \draw (4.25, 0) -- (4.75, 0);
    \draw (5.25, 0) -- (5.75, 0);
    \draw (6.25, 0) -- (6.75, 0);
    \draw (7.25, 0) -- (7.75, 0);
    \draw (8.25, 0) -- (8.75, 0);
    \draw (0.25, -2) -- (0.75, -2);
    \draw (1.25, -2) -- (1.75, -2);
    \draw (2.25, -2) -- (2.75, -2);
    \draw (3.25, -2) -- (3.75, -2);
    \draw (4.25, -2) -- (4.75, -2);
    \draw (5.25, -2) -- (5.75, -2);
    \draw (6.25, -2) -- (6.75, -2);
    \draw (7.25, -2) -- (7.75, -2);
    \draw (8.25, -2) -- (8.75, -2);
    \draw (9.25, -2) -- (9.75, -2);
    \draw (0.25, -4) -- (0.75, -4);
    \draw (1.25, -4) -- (1.75, -4);
    \draw (2.25, -4) -- (2.75, -4);
    \draw (3.25, -4) -- (3.75, -4);
    \draw (4.25, -4) -- (4.75, -4);
    \draw (5.25, -4) -- (5.75, -4);
    \draw (6.25, -4) -- (6.75, -4);
    \draw (7.25, -4) -- (7.75, -4);
    \draw (8.25, -4) -- (8.75, -4);
    \draw (9.25, -4) -- (9.75, -4);
    \draw (0.25, -6) -- (0.75, -6);
    \draw (1.25, -6) -- (1.75, -6);
    \draw (2.25, -6) -- (2.75, -6);
    \draw (3.25, -6) -- (3.75, -6);
    \draw (4.25, -6) -- (4.75, -6);
    \draw (5.25, -6) -- (5.75, -6);
    \draw (6.25, -6) -- (6.75, -6);
    \draw (7.25, -6) -- (7.75, -6);
    \draw (8.25, -6) -- (8.75, -6);
    \draw (9.25, -6) -- (9.75, -6);
    \draw (1.25, -8) -- (1.75, -8);
    \draw (2.25, -8) -- (2.75, -8);
    \draw (3.25, -8) -- (3.75, -8);
    \draw (4.25, -8) -- (4.75, -8);
    \draw (5.25, -8) -- (5.75, -8);
    \draw (6.25, -8) -- (6.75, -8);
    \draw (7.25, -8) -- (7.75, -8);
    \draw (8.25, -8) -- (8.75, -8);
    \draw (9.25, -8) -- (9.75, -8);
    \draw (0, -0.25) -- (0, -0.75);
    \draw (4, -0.25) -- (4, -0.75);
    \draw (8, -0.25) -- (8, -0.75);
    \draw (0, -1.25) -- (0, -1.75);
    \draw (4, -1.25) -- (4, -1.75);
    \draw (8, -1.25) -- (8, -1.75);
    \draw (2, -2.25) -- (2, -2.75);
    \draw (6, -2.25) -- (6, -2.75);
    \draw (10, -2.25) -- (10, -2.75);
    \draw (2, -3.25) -- (2, -3.75);
    \draw (6, -3.25) -- (6, -3.75);
    \draw (10, -3.25) -- (10, -3.75);
    \draw (0, -4.25) -- (0, -4.75);
    \draw (4, -4.25) -- (4, -4.75);
    \draw (8, -4.25) -- (8, -4.75);
    \draw (0, -5.25) -- (0, -5.75);
    \draw (4, -5.25) -- (4, -5.75);
    \draw (8, -5.25) -- (8, -5.75);
    \draw (2, -6.25) -- (2, -6.75);
    \draw (6, -6.25) -- (6, -6.75);
    \draw (10, -6.25) -- (10, -6.75);
    \draw (2, -7.25) -- (2, -7.75);
    \draw (6, -7.25) -- (6, -7.75);
    \draw (10, -7.25) -- (10, -7.75);
    \end{tikzpicture}
    \caption{Qubit topology of the \textit{ibmq\_brooklyn} quantum processor.  The connecting lines between qubits indicate the qubit pairs for which the $CX$ gate is supported at the hardware level.  The qubits used in the implementations are shaded grey.}
    \label{fig:brooklyn}
\end{figure*}

\begin{table}
\centering
\begin{minipage}{11.5cm}
\centering
\begin{tabular}{|l|l|l|l|l|}
\hline
\textbf{Qubit} & $\boldsymbol{T_1}\ \boldsymbol{(}\boldsymbol{\mu}\mathbf{s}\boldsymbol{)}$ & $\boldsymbol{T_2}\ \boldsymbol{(}\boldsymbol{\mu}\mathbf{s}\boldsymbol{)}$ & $\boldsymbol{\sqrt{X}}$ \textbf{Error} & \textbf{Readout Error} \\ \hline
6              & $\,\,\,$71.40$\pm$10.65  & $\,\,\,$93.79$\pm$14.82  & 0.000592$\pm$0.000205     & 0.0389$\pm$0.0185      \\ \hline
5              & $\,\,\,$86.52$\pm$18.82  & 104.75$\pm$22.24 & 0.000278$\pm$0.000071     & 0.0220$\pm$0.0029      \\ \hline
4              & $\,\,\,$81.47$\pm$12.85  & $\,\,\,$83.96$\pm$13.72  & 0.000370$\pm$0.000186     & 0.0219$\pm$0.0060      \\ \hline
3              & $\,\,\,$78.78$\pm$19.86  & $\,\,\,$97.25$\pm$25.10  & 0.000344$\pm$0.000108     & 0.0218$\pm$0.0074      \\ \hline
2              & $\,\,\,$74.44$\pm$16.09  & $\,\,\,$74.86$\pm$12.80  & 0.000561$\pm$0.000511     & 0.0285$\pm$0.0164      \\ \hline
1              & $\,\,\,$87.94$\pm$14.94  & 101.13$\pm$17.78 & 0.000698$\pm$0.000916     & 0.0384$\pm$0.0322      \\ \hline
0              & 100.62$\pm$20.16 & 126.49$\pm$23.14 & 0.000331$\pm$0.000157     & 0.0216$\pm$0.0093      \\ \hline
10             & $\,\,\,$78.63$\pm$11.96  & $\,\,\,$92.33$\pm$18.00  & 0.000437$\pm$0.000503     & 0.0256$\pm$0.0227      \\ \hline
13             & $\,\,\,$68.27$\pm$12.25  & $\,\,\,$12.48$\pm$00.00  & 0.000439$\pm$0.000100     & 0.0186$\pm$0.0048      \\ \hline
14             & $\,\,\,$70.60$\pm$14.50  & $\,\,\,$67.43$\pm$13.69  & 0.000725$\pm$0.000933     & 0.0231$\pm$0.0098      \\ \hline
15             & $\,\,\,$82.73$\pm$15.47  & $\,\,\,$74.37$\pm$16.32  & 0.000526$\pm$0.000205     & 0.0305$\pm$0.0178      \\ \hline
24             & $\,\,\,$72.06$\pm$10.64  & $\,\,\,$83.89$\pm$13.06  & 0.000408$\pm$0.000104     & 0.0173$\pm$0.0039      \\ \hline
29             & $\,\,\,$73.86$\pm$10.95  & $\,\,\,$68.66$\pm$08.13  & 0.000434$\pm$0.000071     & 0.0208$\pm$0.0052      \\ \hline
30             & $\,\,\,$85.58$\pm$15.80  & $\,\,\,$28.83$\pm$02.31  & 0.000344$\pm$0.000288     & 0.0219$\pm$0.0036      \\ \hline
31             & $\,\,\,$71.68$\pm$13.27  & $\,\,\,$81.86$\pm$14.98  & 0.000339$\pm$0.000053     & 0.0164$\pm$0.0022      \\ \hline
39             & $\,\,\,$61.83$\pm$07.62  & $\,\,\,$78.85$\pm$10.48  & 0.000378$\pm$0.000092     & 0.0231$\pm$0.0083      \\ \hline
45             & $\,\,\,$75.02$\pm$11.88  & $\,\,\,$45.20$\pm$06.90  & 0.000352$\pm$0.000079     & 0.0237$\pm$0.0045      \\ \hline
46             & $\,\,\,$65.20$\pm$09.15  & $\,\,\,$79.14$\pm$14.64  & 0.000448$\pm$0.000155     & 0.0211$\pm$0.0054      \\ \hline
47             & $\,\,\,$72.76$\pm$13.58  & $\,\,\,$78.98$\pm$15.52  & 0.000366$\pm$0.000065     & 0.0232$\pm$0.0056      \\ \hline
48             & $\,\,\,$70.91$\pm$11.21  & $\,\,\,$77.97$\pm$13.41  & 0.000503$\pm$0.000350     & 0.0319$\pm$0.0214      \\ \hline
49             & $\,\,\,$68.51$\pm$15.99  & $\,\,\,$79.81$\pm$16.21  & 0.000477$\pm$0.000765     & 0.0274$\pm$0.0304      \\ \hline
50             & $\,\,\,$77.36$\pm$13.30  & $\,\,\,$72.40$\pm$11.81  & 0.000329$\pm$0.000045     & 0.0145$\pm$0.0026      \\ \hline
51             & $\,\,\,$66.58$\pm$10.78  & $\,\,\,$78.75$\pm$13.68  & 0.000447$\pm$0.000174     & 0.0260$\pm$0.0123      \\ \hline
54             & $\,\,\,$78.62$\pm$12.69  & $\,\,\,$81.20$\pm$12.27  & 0.000302$\pm$0.000068     & 0.0168$\pm$0.0025      \\ \hline
64             & $\,\,\,$44.31$\pm$11.38  & $\,\,\,$46.44$\pm$17.45  & 0.000765$\pm$0.000585     & 0.0451$\pm$0.0325      \\ \hline
63             & $\,\,\,$67.51$\pm$15.56  & $\,\,\,$67.97$\pm$12.16  & 0.000419$\pm$0.000176     & 0.0306$\pm$0.0114      \\ \hline
62             & $\,\,\,$80.82$\pm$21.53  & $\,\,\,$88.21$\pm$15.81  & 0.000517$\pm$0.000641     & 0.0358$\pm$0.0295      \\ \hline
61             & $\,\,\,$74.77$\pm$12.04  & $\,\,\,$90.44$\pm$19.60  & 0.000387$\pm$0.000231     & 0.0266$\pm$0.0074      \\ \hline
60             & $\,\,\,$67.55$\pm$14.51  & $\,\,\,$78.66$\pm$16.35  & 0.000471$\pm$0.000259     & 0.0228$\pm$0.0049      \\ \hline
59             & $\,\,\,$76.66$\pm$16.90  & $\,\,\,$90.30$\pm$22.01  & 0.000600$\pm$0.000495     & 0.0280$\pm$0.0178      \\ \hline
58             & $\,\,\,$71.75$\pm$13.95  & $\,\,\,$81.18$\pm$10.42  & 0.000900$\pm$0.000398     & 0.0524$\pm$0.0314      \\ \hline
\end{tabular}
\end{minipage}%
\begin{minipage}{5.5cm}
\centering
\begin{tabular}{|l|l|}
\hline
\textbf{Qubit Pair} & $\boldsymbol{C}\boldsymbol{X}$ \textbf{Error} \\ \hline
6--5                & 0.00951$\pm$0.00206 \\ \hline
5--4                & 0.00743$\pm$0.00243 \\ \hline
4--3                & 0.00893$\pm$0.00609 \\ \hline
3--2                & 0.01202$\pm$0.00332 \\ \hline
2--1                & 0.01803$\pm$0.01495 \\ \hline
1--0                & 0.01559$\pm$0.01385 \\ \hline
0--10               & 0.00913$\pm$0.00448 \\ \hline
10--13              & 0.01282$\pm$0.00480 \\ \hline
13--14              & 0.01490$\pm$0.00824 \\ \hline
14--15              & 0.01438$\pm$0.00651 \\ \hline
15--24              & 0.01101$\pm$0.00295 \\ \hline
24--29              & 0.01207$\pm$0.00398 \\ \hline
29--30              & 0.01093$\pm$0.00331 \\ \hline
30--31              & 0.00928$\pm$0.00338 \\ \hline
31--39              & 0.00952$\pm$0.00183 \\ \hline
39--45              & 0.00972$\pm$0.00222 \\ \hline
45--46              & 0.01035$\pm$0.00234 \\ \hline
46--47              & 0.01126$\pm$0.00148 \\ \hline
47--48              & 0.01103$\pm$0.00537 \\ \hline
48--49              & 0.01393$\pm$0.01103 \\ \hline
49--50              & 0.01242$\pm$0.00679 \\ \hline
50--51              & 0.00788$\pm$0.00112 \\ \hline
51--54              & 0.01040$\pm$0.00134 \\ \hline
54--64              & 0.01586$\pm$0.00855 \\ \hline
64--63              & 0.01822$\pm$0.00534 \\ \hline
63--62              & 0.01206$\pm$0.01430 \\ \hline
62--61              & 0.01245$\pm$0.00374 \\ \hline
61--60              & 0.01291$\pm$0.00595 \\ \hline
60--59              & 0.01098$\pm$0.00438 \\ \hline
59--58              & 0.01780$\pm$0.00942 \\ \hline
\end{tabular}
\end{minipage}
\caption{Calibration information for the \textit{ibmq\_brooklyn} quantum processor averaged over the time period during which circuits were run and data was obtained.  The single-qubit calibration information for the relevant qubits is shown on the left.  $T_1$ and $T_2$ are the relaxation and dephasing times respectively of the qubits.  The $CX$ error rates for relevant qubit pairs are shown on the right.}
\label{tab:calibrationbrooklyn}
\end{table}

\section{Process tomography}\label{append:channeltomo} 

For a given measurement-based gate, quantum process tomography was performed on each of the three sets of qubits on which that gate was implemented in the interleaved sequences, that were implemented to obtain the data required to estimate the fidelity of the given measurement-based gate.  To this end, we employed the quantum process tomography method proposed for single-qubit processes by Nielsen and Chuang~\cite{QPT1, QPT2}.  This single-qubit process tomography method was reviewed extensively in our previous work~\cite{previous}.  The method amounts to performing quantum state tomography to infer the output state for four different probe input states, namely $\ket{0}$, $\ket{1}$, $\ket{+}=\left(\ket{0}+\ket{1}\right)/\sqrt{2}$ and $\ket{+_{y}}=\left(\ket{0}+\text{i}\ket{1}\right)/\sqrt{2}$.  The state tomography results are then used to construct a 4 by 4 process matrix $\chi$, which completely characterises a given implementation of a single-qubit process.

To perform quantum process tomography for one of the three implementations of a measurement-based gate, on one of the three sets of qubits used in the interleaved sequences, the required linear cluster state was prepared on the set of qubits (with a given probe input state prepared on the first qubit), single-qubit measurements were performed on all but the final qubit, and quantum state tomography was performed on the final qubit to infer the output state, for each byproduct, for the given probe input state.  Each of the twelve circuits needed for process tomography (three circuits for state tomography to infer the output state for each of the four probe input states) was run with 8192 shots on the relevant processor.  For each byproduct, matrix multiplication was once again used to apply the appropriate Pauli correction to the density matrices, constructed by performing state tomography for each of the four probe input states.  The Pauli corrected density matrices were used to construct a process matrix for each byproduct.  An average process matrix for a given measurement-based implementation was then calculated from the process matrices obtained for the different byproducts.

The quantum process fidelity of one of the three implementations of a measurement-based gate $G$, on one of the three sets of qubits used in the interleaved sequences, was calculated using
\begin{equation}
F_{G}^{P}(\chi,\widetilde{\chi})=\text{Tr}\left(\sqrt{\sqrt{\chi}\widetilde{\chi}\sqrt{\chi}}\right),
\end{equation}
where $\chi$ is the process matrix for the ideal implementation of $G$ and $\widetilde{\chi}$ is the average process matrix obtained by performing quantum process tomography for the given implementation of $G$.  The Haar-averaged gate fidelity of each of the three implementations of $G$ was then calculated using
\begin{equation}
F_{G}=\frac{d\left(F_{G}^{P}(\chi,\widetilde{\chi})\right)^2+1}{d+1},
\end{equation}
where $d=2$ for single qubits~\cite{SRB3}.  Based on the fidelities calculated from process tomography results in our previous work~\cite{previous}, we expect the uncertainty in the fidelity of a given implementation of $G$ to be small compared to the range of fidelities for the three different implementations of $G$ on the three different sets of qubits.  We therefore determined only a single fidelity for each implementation of $G$, with no estimated uncertainty, to avoid unnecessary use of processor time.

\section{Qubits used for fidelity estimation of noisier gates}\label{append:qubitsnoisier} 

To perform fidelity estimation for the noisier measurement-based implementations, we implemented reference sequences with lengths $m\in\{1,2,3\}$ and interleaved sequences with $m\in\{1,2,3\}$, for the 4-qubit and 6-qubit Hadamard gate as well as the 5-qubit and 7-qubit $T$ gate, on the \textit{ibmq\_brooklyn} quantum processor.  The qubit topology of the \textit{ibmq\_brooklyn} quantum processor is shown in \figref{fig:brooklyn}, with the qubits used in the implementations shaded grey.  These qubits were once again chosen for their low error rates compared to other qubits.  For each $m$ (reference or interleaved), the required linear cluster state was prepared along the length of shaded qubits, starting at qubit 6.  For example, to implement the interleaved sequence for the 5-qubit $T$ gate with $m=1$, the required 9-qubit linear cluster state was prepared on the qubits 6, 5, 4, 3, 2, 1, 0, 10 and 13, single-qubit measurements were performed on the qubits 6, 5, 4, 3, 2, 1, 0 and 10, and quantum state tomography was performed on qubit 13.  All thirty-one shaded qubits, from qubit 6 through to qubit 58, were used to implement the interleaved sequence for the 7-qubit $T$ gate with $m=3$, which required a 31-qubit linear cluster state.

The required data was obtained over a period of 86 days on the \textit{ibmq\_brooklyn} quantum processor.  On each day on which circuits were run, the relevant calibration information was recorded at 19:00 GMT.  \tabref{tab:calibrationbrooklyn} shows the average of the calibration information recorded during this period, with the uncertainties given by the standard deviation.


\begin{thebibliography}{99}

\bibitem{application1} Shor, P. W. Polynomial-time algorithms for prime factorization and discrete logarithms on a quantum computer. \textit{SIAM J. Sci. Comput.} \textbf{26}, 1484--1509 (1997).

\bibitem{application2} Grover, L. K. Quantum mechanics helps in searching for a needle in a haystack. \textit{Phys. Rev. Lett.} \textbf{79}, 325--328 (1997).

\bibitem{application3} Feynman, R. P. Simulating physics with computers. \textit{Int. J. Theor. Phys.} \textbf{21}, 467--488 (1982).

\bibitem{application4} Biamonte, J., Wittek, P., Pancotti, N., Rebentrost, P., Wiebe, N. \& Lloyd, S. Quantum machine learning. \textit{Nature} \textbf{549}, 195--202 (2017).

\bibitem{application5} Hauke, P., Katzgraber, H. G., Lechner, W., Nishimori, H. \& Oliver, W. D. Perspectives of quantum annealing: methods and implementations. \textit{Rep. Prog. Phys.} \textbf{83}, 054401 (2020).

\bibitem{QCM1} Deutsch, D. E. Quantum computational networks. \textit{Proc. Math. Phys. Eng. Sci.} \textbf{425}, 73--90 (1989).

\bibitem{QCM2} Barenco, A., Bennett, C. H., Cleve, R., DiVincenzo, D. P., Margolus, N., Shor, P., Sleator, T., Smolin, J. \& Weinfurter, H. Elementary gates for quantum computation. \textit{Phys. Rev.}~A \textbf{52}, 3457--3467 (1995).

\bibitem{MBQC1} Raussendorf, R. \& Briegel, H. J. A one-way quantum computer. \textit{Phys. Rev. Lett.} \textbf{86}, 5188--5191 (2001).

\bibitem{MBQC2} Raussendorf, R. \& Briegel, H. J. Computational model underlying the one-way quantum computer. \textit{Quantum Inf. Comput.} \textbf{2}, 443--486 (2002).

\bibitem{MBQC3} Raussendorf, R., Browne, D. E. \& Briegel, H. J. Measurement-based quantum computation on cluster states. \textit{Phys. Rev.}~A \textbf{68}, 022312 (2003).

\bibitem{MBQC4} Briegel, H. J., Browne, D. E., D\"{u}r, W., Raussendorf, R. \& Van der Nest, M. Measurement-based quantum computation. \textit{Nat. Phys.} \textbf{5}, 19--26 (2009).

\bibitem{MBQCexam1} Nielsen, M. A. Optical quantum computation using cluster states. \textit{Phys. Rev. Lett.} \textbf{93}, 040503 (2004).

\bibitem{MBQCexam2} Browne, D. E. \& Rudolph, T. Resource-efficient linear optical quantum computation. \textit{Phys. Rev. Lett.} \textbf{95}, 010501 (2005).

\bibitem{MBQCexam3} Walther, P., Resch, K. J., Rudolph, T., Schenck, E., Weinfurter, H., Vedral, V., Aspelmeyer, M. \& Zeilinger, A. Experimental one-way quantum computing. \textit{Nature} \textbf{434}, 169--176 (2005).

\bibitem{MBQCexam4} Tanamoto, T., Liu, Y. X., Fujita, S., Hu, X. \& Nori, F. Producing cluster states in charge qubits and flux qubits. \textit{Phys. Rev. Lett.} \textbf{97}, 230501 (2006).

\bibitem{MBQCexam5} Vaucher, B., Nunnenkamp, A. \& Jaksch, D. Creation of resilient entangled states and a resource for measurement-based quantum computation with optical superlattices. \textit{New J. Phys.} \textbf{10}, 023005 (2008).

\bibitem{MBQCexam6} Weinstein, Y. S., Hellberg, C. S. \& Levy, J. Quantum-dot cluster-state computing with encoded qubits. \textit{Phys. Rev.}~A \textbf{72}, 020304(R) (2005).

\bibitem{noise1} Preskill, J. Quantum computing in the NISQ era and beyond. \textit{Quantum} \textbf{2}, 79 (2018).

\bibitem{noise2} Georgopoulos, K., Emary, C. \& Zuliani, P. Modelling and simulating the noisy behaviour of near-term quantum computers. \textit{Phys. Rev.}~A \textbf{104}, 062432 (2021).

\bibitem{noise3} Skosana, U. \& Tame, M. S. Demonstration of Shor's factoring algorithm for $N = 21$ on IBM quantum processors. \textit{Sci. Rep.} \textbf{11}, 16599 (2021).

\bibitem{QPT1} Nielsen, M. A. \& Chuang, I. L. Quantum process tomography in \textit{Quantum Computation and Quantum Information: 10th Anniversary Edition} 389--394 (Cambridge University Press, 2010).

\bibitem{QPT2} Chuang, I. L. \& Nielsen, M. A. Prescription for experimental determination of the dynamics of a quantum black box. \textit{J. Mod. Opt.} \textbf{44}, 2455--2467 (1997).

\bibitem{QPT3} Poyatos, J. F., Cirac, J. I. \& Zoller, P. Complete characterization of a quantum process: the two-bit quantum gate. \textit{Phys. Rev. Lett.} \textbf{78}, 390--393 (1997).

\bibitem{fidelity} Sanders, Y. R., Wallman, J. J. \& Sanders, B. C. Bounding quantum gate error rate based on reported average fidelity. \textit{New J. Phys.} \textbf{18}, 012002 (2016).

\bibitem{RBthesis} Boone, K. Concepts and methods for benchmarking quantum computers. (University of Waterloo, 2021).

\bibitem{SRB1} Emerson, J., Alicki, R. \& \.{Z}yczkowski, K. Scalable noise estimation with random unitary operators. \textit{J. opt., B Quantum semiclass. opt.} \textbf{7}, 347--352 (2005).

\bibitem{SRB2} L\'{e}vi, B., L\'{o}pez, C. C., Emerson, J. \& Cory, D. G. Efficient error characterization in quantum information processing. \textit{Phys. Rev.}~A \textbf{75}, 022314 (2007).

\bibitem{SRB3} Dankert, C., Cleve, R., Emerson, J. \& Livine, E. Exact and approximate unitary 2-designs and their application to fidelity estimation. \textit{Phys. Rev.}~A \textbf{80}, 012304 (2009).

\bibitem{SRB4} Magesan, E., Gambetta, J. M. \& Emerson, J. Scalable and robust randomized benchmarking of quantum processes. \textit{Phys. Rev. Lett.} \textbf{106}, 180504 (2011).

\bibitem{SRB5} Magesan, E., Gambetta, J. M. \& Emerson, J. Characterizing quantum gates via randomized benchmarking. \textit{Phys. Rev.}~A \textbf{85}, 042311 (2012).

\bibitem{SRB6} Wallman, J. J. Randomized benchmarking with gate-dependent noise. \textit{Quantum} \textbf{2}, 47 (2018).

\bibitem{SRB7} Proctor, T., Rudinger, K., Young, K., Sarovar, M. \& Blume-Kohout, R. What randomized benchmarking actually measures. \textit{Phys. Rev. Lett.} \textbf{119}, 130502 (2017).

\bibitem{SRB8} Carignan-Dugas, A., Boone, K., Wallman, J. J. \& Emerson, J. From randomized benchmarking experiments to gate-set circuit fidelity: how to interpret randomized benchmarking decay parameters. \textit{New J. Phys.} \textbf{20}, 092001 (2018).

\bibitem{SRB9} Knill, E., Leibfried, D., Reichle, R., Britton, J., Blakestad, R. B., Jost, J. D., Langer, C., Ozeri, R., Seidelin, S. \& Wineland, D. J. Randomized benchmarking of quantum gates. \textit{Phys. Rev.}~A \textbf{77}, 012307 (2008).

\bibitem{SRB10} Boone, K., Carignan-Dugas, A., Wallman, J. J. \& Emerson, J. Randomized benchmarking under different gatesets. \textit{Phys. Rev.}~A \textbf{99}, 032329 (2019).

\bibitem{SRB11} Brown, W. G. \& Eastin, B. Randomized benchmarking with restricted gate sets. \textit{Phys. Rev.}~A \textbf{97}, 062323 (2018).

\bibitem{SRB12} Hashagen, A. K., Flammia, S. T., Gross, D. \& Wallman, J. J. Real randomized benchmarking. \textit{Quantum} \textbf{2}, 85 (2018).

\bibitem{IRB1} Magesan, E., Gambetta, J. M., Johnson, B. R., Ryan, C. A., Chow, J. M., Merkel, S. T., Da Silva, M. P., Keefe, G. A., Rothwell, M. B., Ohki, T. A., Ketchen, M. B. \& Steffen, M. Efficient measurement of quantum gate error by interleaved randomized benchmarking. \textit{Phys. Rev. Lett.} \textbf{109}, 080505 (2012).

\bibitem{IRB2} Gaebler, J. P., Meier, A. M., Tan, T. R., Bowler, R., Lin, Y., Hanneke, D., Jost, J. D., Home, J. P., Knill, E., Leibfried, D. \& Wineland, D. J. Randomized benchmarking of multiqubit gates. \textit{Phys. Rev. Lett.} \textbf{108}, 260503 (2012).

\bibitem{IRB3} Carignan-Dugas, A., Wallman, J. J. \& Emerson, J. Bounding the average gate fidelity of composite channels using the unitarity. \textit{New J. Phys.} \textbf{21}, 053016 (2019).

\bibitem{IRB4} Carignan-Dugas, A., Wallman, J. J. \& Emerson, J. Characterizing universal gate sets via dihedral benchmarking. \textit{Phys. Rev.}~A \textbf{92}, 060302(R) (2015).

\bibitem{IRB5} Harper, R. \& Flammia, S. T. Estimating the fidelity of $T$ gates using standard interleaved randomized benchmarking. \textit{Quantum Sci. Technol.} \textbf{2}, 015008 (2017).

\bibitem{RBtrappedions1} Ballance, C. J., Harty, T. P., Linke, N. M., Sepiol, M. A. \& Lucas, D. M. High-fidelity quantum logic gates using trapped-ion hyperfine qubits. \textit{Phys. Rev. Lett.} \textbf{117}, 060504 (2016).

\bibitem{RBsuperconducting1} Chow, J. M., DiCarlo, L., Gambetta, J. M., Motzoi, F., Frunzio, L., Girvin, S. M. \& Schoelkopf, R. J. Implementing optimal control pulse shaping for improved single-qubit gates. \textit{Phys. Rev.}~A \textbf{82}, 040305(R) (2010).

\bibitem{RBsuperconducting2} C\'{o}rcoles, A. D., Gambetta, J. M., Chow, J. M., Smolin, J. A., Ware, M., Strand, J., Plourde, B. L. T. \& Steffen, M. Process verification of two-qubit quantum gates by randomized benchmarking. \textit{Phys. Rev.}~A \textbf{87}, 030301(R) (2013).

\bibitem{RBsuperconducting3} Barends, R., Kelly, J., Megrant, A., Veitia, A., Sank, D., Jeffrey, E., White, T. C., Mutus, J., Fowler, A. J., Campbell, B., Chen, Y., Chen, Z., Chiaro, B., Dunsworth, A., Neill, C., O'Malley, P., Roushan, P., Vainsencher, A., Wenner, J., Korotkov, A. N., Cleland, A. N. \& Martinis, J. M. Logic gates at the surface code threshold: superconducting qubits poised for fault-tolerant quantum computing. \textit{Nature} \textbf{508}, 500--503 (2014).

\bibitem{RBsuperconducting4} McKay, D. C., Sheldon, S., Smolin, J. A., Chow, J. M. \& Gambetta, J. M. Three-qubit randomized benchmarking. \textit{Phys. Rev. Lett.} \textbf{122}, 200502 (2019).

\bibitem{RBNMR1} Ryan, C. A., Laforest, M. \& Laflamme, R. Randomized benchmarking of single- and multi-qubit control in liquid-state NMR quantum information processing. \textit{New J. Phys.} \textbf{11}, 013034 (2009).

\bibitem{RBcoldatoms1} Olmschenk, S., Chicireanu, R., Nelson, K. D. \& Porto, J. V. Randomized benchmarking of atomic qubits in an optical lattice. \textit{New J. Phys.} \textbf{12}, 113007 (2010).

\bibitem{RBcoldatoms2} Xia, T., Lichtman, M., Maller, K., Carr, A. W., Piotrowicz, M. J., Isenhower, L. \& Saffman, M. Randomized benchmarking of single-qubit gates in a 2D array of neutral-atom qubits. \textit{Phys. Rev. Lett.} \textbf{114}, 100503 (2015).

\bibitem{RBqdot1} Veldhorst, M., Hwang, J. C. C., Yang, C. H., Leenstra, A. W., De Ronde, B., Dehollain, J. P., Muhonen, J. T., Hudson, F. E., Itoh, K. M., Morello, A. \& Dzurak, A. S. An addressable quantum dot qubit with fault-tolerant control-fidelity. \textit{Nat. Nanotechnol.} \textbf{9}, 981--985 (2014).

\bibitem{MRB} Mayer, K., Hall, A., Gatterman, T., Halit, S. K., Lee, K., Bohnet, J., Gresh, D., Hankin, A., Gilmore, K. \& Gaebler, J. Theory of mirror benchmarking and demonstration on a quantum computer. arXiv:2108.10431 (2021).

\bibitem{HRB} Nakata, Y., Zhao, D., Okuda, T., Bannai, E., Suzuki, Y., Tamiya, S., Heya, K., Yan, Z., Zuo, K., Tamate, S., Tabuchi, Y. \& Nakamura, Y. Quantum circuits for exact unitary $t$-designs and applications to higher-order randomized benchmarking. \textit{PRX Quantum} \textbf{2}, 030339 (2021).

\bibitem{RCS} Liu, Y., Otten, M., Bassirianjahromi, R., Jiang, L. \& Fefferman, B. Benchmarking near-term quantum computers via random circuit sampling. arXiv:2105.05232 (2021).

\bibitem{MBSRB} Alexander, R. N., Turner, P. S. \& Bartlett, S. D. Randomized benchmarking in measurement-based quantum computing. \textit{Phys. Rev.}~A \textbf{94}, 032303 (2016).

\bibitem{previous} Strydom, C. \& Tame, M. S. Implementation of single-qubit measurement-based $t$-designs using IBM processors. \textit{Sci. Rep.} \textbf{12}, 5014 (2022).

\bibitem{IBM} IBM Quantum Experience, https://quantum-computing.ibm.com/. Accessed 22 February 2022.

\bibitem{MBQCIBM} Yang, Z. P., Ku, H. Y., Baishya, A., Zhang, Y. R., Kockum, A. F., Chen, Y. N., Li, F. L., Tsai, J. S. \& Nori, F. Deterministic one-way logic gates on a cloud quantum computer. \textit{Phys. Rev.}~A \textbf{105}, 042610 (2022).

\bibitem{HandT} Nielsen, M. A. \& Chuang, I. L. Universal quantum gates in \textit{Quantum Computation and Quantum Information: 10th Anniversary Edition} 188--202 (Cambridge University Press, 2010).

\bibitem{MBQCft} Raussendorf, R., Harrington, J. \& Goyal, K. A fault-tolerant one-way quantum computer. \textit{Ann. Phys.} \textbf{321}, 2242--2270 (2006).

\bibitem{MBQClc} Nielsen, M. A. Cluster-state quantum computation. \textit{Rep. Math. Phys.} \textbf{57}, 147--161 (2006).

\bibitem{MBT1} Turner, P. S. \& Markham, D. Derandomising quantum circuits with measurement-based unitary designs. \textit{Phys. Rev. Lett.} \textbf{116}, 200501 (2016).

\bibitem{MBT2} Mezher, R., Ghalbouni, J., Dgheim, J. \& Markham, D. Efficient quantum pseudorandomness with simple graph states. \textit{Phys. Rev.}~A \textbf{97}, 022333 (2018).

\bibitem{WCE1} Aliferis, P., Gottesman, D. \& Preskill, J. Quantum accuracy threshold for concatenated distance-3 codes. \textit{Quantum Inf. Comput.} \textbf{6}, 97--165 (2006).

\bibitem{WCE2} Wallman, J. J. \& Flammia, S. T. Randomized benchmarking with confidence. \textit{New J. Phys.} \textbf{16}, 103032 (2014).

\bibitem{WCE3} Wallman, J. J. Bounding experimental quantum error rates relative to fault-tolerant thresholds. arXiv:1511.00727 (2015).

\bibitem{WCE4} Kueng, R., Long, D. M., Doherty, A. C. \& Flammia, S. T. Comparing experiments to the fault-tolerance threshold. \textit{Phys. Rev. Lett.} \textbf{117}, 170502 (2016).

\bibitem{Clifford} Gottesman, D. Theory of fault-tolerant quantum computation. \textit{Phys. Rev.}~A \textbf{57}, 127--137 (1998).

\bibitem{IRBgeneral} Kimmel, S., Da Silva, M. P., Ryan, C. A., Johnson, B. R. \& Ohki, T. Robust extraction of tomographic information via randomized benchmarking. \textit{Phys. Rev.}~X \textbf{4}, 011050 (2014).

\bibitem{GKtheorem} Gottesman, D. The Heisenberg representation of quantum computers in \textit{Group 22: Proceedings of the XXII International Colloquium on Group Theoretical Methods in Physics} 32--43 (International Press, 1999).

\bibitem{previousnoise} Strydom, C. \& Tame, M. S. Investigating the effect of noise channels on the quality of unitary $t$-designs. arXiv:2203.13771 (2022).

\bibitem{RBrep} Granade, C., Ferrie, C. \& Cory, D. G. Accelerated randomized benchmarking. \textit{New J. Phys.} \textbf{17}, 013042 (2015).

\bibitem{lcspreparation} Mooney, G. J., Hill, C. D. \& Hollenberg, L. C. L. Entanglement in a 20-qubit superconducting quantum computer. \textit{Sci. Rep.} \textbf{9}, 13465 (2019).

\bibitem{tomography} StateTomographyFitter, https://qiskit.org/documentation/ stubs/qiskit.ignis.verification.StateTomographyFitter.html. Accessed 22 February 2022.

\bibitem{photonic1} Adcock, J. C., Vigliar, C., Santagati, R., Silverstone, J. W. \& Thompson, M. G. Programmable four-photon graph states on a silicon chip. \textit{Nat. Commun.} \textbf{10}, 3528 (2019).

\bibitem{photonic2} Wang, J., Sciarrino, F., Laing, A. \& Thompson, M. G. Integrated photonic quantum technologies. \textit{Nat. Photonics} \textbf{14}, 273--284 (2020).

\bibitem{photonic3} Bartolucci, S., Birchall, P., Bombin, H., Cable, H., Dawson, C., Gimeno-Segovia, M., Johnston, E., Kieling, K., Nickerson, N., Pant, M., Pastawski, F., Rudolph, T. \& Sparrow, C. Fusion-based quantum computation. arXiv:2101.09310 (2021).

\bibitem{photonic4} Bartolucci, S., Birchall, P., Bonneau, D., Cable, H., Gimeno-Segovia, M., Kieling, K., Nickerson, N., Rudolph, T. \& Sparrow, C. Switch networks for photonic fusion-based quantum computing. arXiv:2109.13760 (2021).

\end{thebibliography}
\end{document}